\documentclass[shortauth]{aa}  %,onecolumn
\usepackage{graphicx}
\usepackage{caption, subcaption}
\usepackage{xcolor}
\usepackage[switch]{lineno}
\usepackage{txfonts}
\usepackage{pifont}
\usepackage{bm}
\usepackage{amsmath,array,graphicx}

\usepackage[hidelinks]{hyperref}
\usepackage[normalem]{ulem}
\usepackage{orcidlink}
\usepackage{booktabs}

\begin{document} 

   \title{The impact of outliers on pulsar timing arrays}
\author{ Giulia Fumagalli\orcidlink{0009-0004-2044-989X}\inst{1,2}, Golam Shaifullah\orcidlink{0000-0002-8452-4834}\inst{1,2,3}, Alberto Sesana\orcidlink{0000-0003-4961-1606}\inst{1,2,4}}
\institute{
    Dipartimento di Fisica ``G. Occhialini", Universit{\'a} degli Studi di Milano-Bicocca, Piazza della Scienza 3, I-20126 Milano, Italy\\
    \email{g.fumagalli47@campus.unimib.it}
    \and
    INFN, Sezione di Milano-Bicocca, Piazza della Scienza 3, I-20126 Milano, Italy
    \and
    INAF - Osservatorio Astronomico di Cagliari, via della Scienza 5, 09047 Selargius (CA), Italy
    \and
    INAF - Osservatorio Astronomico di Brera, via Brera 20, I-20121 Milano, Italy
     }

\titlerunning{}
\authorrunning{Fumagalli et al}

  \abstract{The detection of gravitational waves with Pulsar Timing Arrays (PTAs)
	requires precise measurement of the difference between the pulsars'
	timing models and their observed pulses, as well as dealing with
	numerous and sometimes hard to diagnose sources of
	noise. Outliers may have an impact on this already difficult
	procedure, especially if the methods used are not robust to such
	anomalous observations. Until now, no complete and practical
	quantification of their effects on PTA data has been provided. With
	this work, we aim to fill this gap. We corrupt  simulated
	datasets featuring an increasing degree of complexity  with varying percentages of uniformly distributed outliers
	and investigate the impact of the latter on the recovery of the
	injected gravitational wave signals and pulsar noise terms. We found that the gravitational waves signal, due
	to its expected correlation, is more robust against these anomalous
	observations when compared to the other injected processes. This
	result is especially relevant in the context of the emerging
	statistical evidence for the gravitational wave background in PTA
	datasets, further strengthening those claims.
	}

   \maketitle
\section{INTRODUCTION} \label{sec:INTRODUCTION}

Supermassive (M$_{BH}>10^8$ M$_{\odot}$) black hole binaries (SMBHBs)
emit nanohertz-frequency gravitational waves (GWs)
during their slow and adiabatic inspiral phase \citep[see e.g.,][and
  references therein]{sesana+2008}. To observe them, it is necessary
to exploit Galactic-scale detectors consisting of arrays of regularly
monitored millisecond pulsars \citep[][]{bkh+82}, whose extreme
rotational stability, leading to their characteristic pulsed
observations, is comparable with the precision of atomic clocks
\citep[see e.g.,][]{Lorimer+2004,hgc+20}. Such detectors are known as
PTAs \citep[][]{fb90}. Detection can be
accomplished by comparing regularly recorded times-of-arrival (ToAs)
of the pulses from each pulsar with theoretical predictions. The
latter derives from a `pulsar timing' model of the pulsars describing
their astrometry, rotational behavior, additional orbital effects if
they are in binary systems, as well as the effects of any intervening
sources of delays such as the interstellar medium (ISM). The outcomes
of this comparison are the pulsar timing \textit{residuals}. As shown
by \citet{Sazhin78,Detweiler79,maggiore+2008} and others, when GWs
cross the space between pulsars and the Earth they perturb the local
space-time along the propagation path of the pulses, inducing a
correlated delay in the timing residuals of each pulsar. This
correlation is a function of the angular separation between pulsar
pairs, and follows the form predicted by \citet[][]{HD+1983}, henceforth,  HD
  correlation.

Recently, four major PTA collaborations, namely the European Timing Array and
Indian Pulsar Timing Arrays \citep[EPTA and InPTA,
  respectively][]{Ferdman+2010,jgp+22}, the North American Nanohertz
Observatory for Gravitational Waves \citep[NANOGrav,][]{nanograv}, the
Parkes Pulsar Timing Array \citep[PPTA,][]{PPTA} and the Chinese PTA
\citep[CPTA][]{cpta}, presented evidence in their data for the presence of a
correlated red noise process that follows the HD correlation. In
addition, MeerTime \citep{bja+20}, the pulsar timing experiment at
MeerKAT -- the expanded Karoo Array Telescope in South Africa -- have
released their first PTA datasets \citep{SA2022}. Together,
all the `regional' PTAs, apart from the CPTA, are combining their datasets
into a common global effort; the International PTA (IPTA)
\citep{Verbist+2016,Perera+2019} to increase the overall sensitivity
of the datasets.

The first gravitational signal that PTAs expect to observe is a
stochastic GW background (GWB), most likely produced by the incoherent
superposition of GWs generated by inspiralling SMBHBs \citep[see
  e.g.,][ and references therein]{Rosado+2015}. Due to the stochastic
nature of this signal, it cannot be included in the deterministic
pulsar timing model and hence it shows up in the residuals. This
effect is relatively weak, and working with an extremely precise
timing model and high-quality data is necessary for successful
detection. This task is even more challenging since the GWB is not the
only contributor to the residuals. As shown in \citet{Chalumeau+2021},
there are also signatures of white (Gaussian or radiometer) noise and
of pulsar intrinsic red noise (RN) or timing noise, which can mask the
GWB. Finally, density fluctuations in the ionized ISM crossing the
line of sight lead to another noise component, quantified as
variations of the pulsar `dispersion measure' (DM). This can be
particularly troublesome since it induces a similar delay in timing
residuals as the GWB. However, the GWB can be distinguished from other
noise sources through its characteristic HD correlation. 

Apart from these competing noise sources, several systematic can
pose challenges to extracting the contribution of the GWB from the
timing residuals. One of these could be \textit{outliers},
pathological observations that can emerge from a process different
from those responsible for most of the data. This kind of observation
can arise from data entry errors, due to recording and measurement
errors, or can be related to rare or unknown astrophysical
events. Regardless of their origin, their presence cannot be ignored,
especially when statistical techniques are applied to the data. The
least-square fitting procedures \citep{rousseeuw+2005} on which pulsar
timing software such as \textsc{Tempo2} \citep{hem06} are based on are
particularly susceptible to the influence of outliers.
As shown in \cite{Vallisneri+vanhaasteren+2017}, the presence of such
anomalous observations can bias the estimation of WN parameters. Some
methods have been proposed in order to take care of outliers
\citep{Vallisneri+vanhaasteren+2017,Wang+2017} in the PTA framework,
although given different processing schemes adopted by different PTAs,
these are yet to become part of standard analysis. Compounding these
issues, purely statistical outliers can easily be conflated by
transient events known to occur in pulsar timing datasets.

In this study, we examine the influence of outliers on the properties
of the recovered common signals in PTA datasets. Specifically, we
search for both, a common \textit{uncorrelated} red noise
(henceforth, CP) process as well as an HD correlated process, in
realistic datasets with noise properties mirroring real data and
including an increasing percentage of uniformly distributed outliers. We also search for such processes in datasets to
which no common process is added. Using Bayesian model selection, we
test for biased recovery when outliers are present in the data.

This investigation is critical, in light of the recent PTA discoveries. In fact, the significance of the reported HD correlated signal ranges between $2\sigma$ and $4\sigma$, thus not yet meeting the {\it golden standard} generally accepted to claim detection. Although consistent with a SMBHB origin, the measured spectral properties of this signal are in mild tension with vanilla models of circular-GW driven SMBHB populations. In fact the data 
favour a background with amplitude pushing towards the upper limit produced by astrophysical models 
\citep{Izquierdo-Villalba+2022} and are 
described by a power-law with a flatter
spectral index than expected from a population of circularized supermassive black hole binaries (SMBHBs). However, uncertainties in the measurements are large and caution should be taken when drawing strong astrophysical conclusions from them \citep[see discussions in][]{eptadr2V}. It is therefore important to assess the robustness of detection and parameter estimation against potential biases arising from the presence of bad data.

The paper is organized as follows.
We describe how we constructed the datasets and their
characteristics in Section \ref{sec:DATASETS AND METHODS}, we present
the results of their analysis in Section \ref{sec:RESULTS}, we interpret those results and discuss their implications for PTA real
data analysis in Section \ref{sec:DISCUSSION}, and summarize our main findings in Section \ref{sec:CONCLUSION}.
\section{DATASETS AND METHODS} \label{sec:DATASETS AND METHODS}

\subsection{Datasets for signal-recovery analysis}\label{sec:datasetsSR}
We generated three PTA-like datasets with an increasing degree of
realism employing \texttt{libstempo} \citep{libstempo}, a python
interface to \textsc{Tempo2} \citep{hem+2006,ehm+2006}. We simulate
ToAs for 25 pulsars observed by the EPTA collaboration, whose data\footnote{available at
\url{https://gitlab.com/IPTA/DR2}} are available in the IPTA second
data release \citep[henceforth the IPTA DR2,][]{Perera+2019}. We chose
to retain the actual starting and ending dates of observations for
each pulsar in the datasets, as well as the number of observations
while varying the cadence to obtain a slightly more uniform yet
irregular distribution of the observations.

The main differences between the three datasets are the number of
observing frequencies and systems, and the way in which we assigned
the values of the WN parameters:

\begin{enumerate}
    \item  \underline{\textit{OneF} dataset}: we assume that for each pulsar, the observations
have been performed at a single observing frequency with a single
telescope;
    \item \underline{\textit{TwoF} dataset}: we introduce two observing frequencies associated with a unique telescope;
    \item \underline{\textit{MultiF} dataset}: we take, for each pulsar, the observing frequencies, observatories, and systems utilized in the actual IPTA DR2 dataset for those pulsars. %
    For some of the analysis we extend this dataset by 10 years, thus producing a \underline{\textit{MultiF+10yr} dataset}.
\end{enumerate}

For the \textit{TwoF} and \textit{MultiF} datasets
we also fitted constant offsets (JUMPs) to account for the use of
multiple systems. To construct all the datasets, we first generated for
each pulsar, idealized ToAs such that, when compared with the pulsar's
timing model, they return zero timing residuals. We assigned realistic
uncertainties $\sigma_{\rm ToA}$ to each observation and then, using our
knowledge from real data analysis \citep[see e.g.,][]{Chalumeau+2021},
we injected white (or Gaussian) noise by rescaling them as follows:
\begin{equation}
   \sigma=\sqrt{EFAC^2\sigma_{\rm ToA}^2+ EQUAD^2}.
\end{equation}
Here EFAC accounts for factorial imperfections in the white-noise
quantification, whereas EQUAD accounts for potential additive sources
of noise that are not naturally included in the formal ToA
uncertainties $\sigma_{\rm ToA}$. The values of EFAC and EQUAD used
change based on the dataset as reported in Table \ref{tab:efac_eqaud}.
\begin{table}
	\centering
	\caption{\small{EFAC and EQUAD used in the simulated
            datasets. TNEF and TNEQ refers to the corresponding values
            reported in the parameters file of the pulsars
            considered.}}
		\begin{tabular}{cll}
    	\hline dataset& EFAC &EQUAD \\ \hline \textit{OneF}& $1$,
        global &$ 10^{-6}$, global\\ \textit{TwoF}& $1$, per system &$
        10^{-6}$, per system\\ \textit{ MultiF}& TNEF, per system &
        TNEQ, per system\\ \hline
     
	\end{tabular}
	\label{tab:efac_eqaud}
\end{table}
For each pulsar, we injected timing noise, which consists of RN
that can be modelled with a power-law power spectral density (PSD)
function of the form:%
\begin{equation}\label{powerlaw}
\mathcal{P}_{RN}(f)=\frac{A_{RN}^2}{12 \pi^2}\left(\frac{f}{\mbox{
    yr}^{-1}}\right)^{-\gamma_{RN}}
\end{equation}
where $A_{RN}$ is the RN amplitude and $\gamma_{RN}$ is the spectral
index. For this signal, we fix the number of Fourier modes to 30,
following \cite{Chen+2021}. The values of the amplitude chosen for
this work have been taken from the single pulsar noise analysis
performed in
\cite{Antoniadis+2022}.%

In the \textit{MultiF} dataset only, we also injected a chromatic
DM noise, which spectrum,  specific for each pulsar,
can be modelled exactly as the RN. We chose the amplitude and a spectral index, again following the analysis carried out in
\cite{Antoniadis+2022}. Following \cite{Chen+2021}, we choose to use
100 Fourier modes to describe this signal. %

\begin{figure}[h]
    \includegraphics[width=\linewidth]{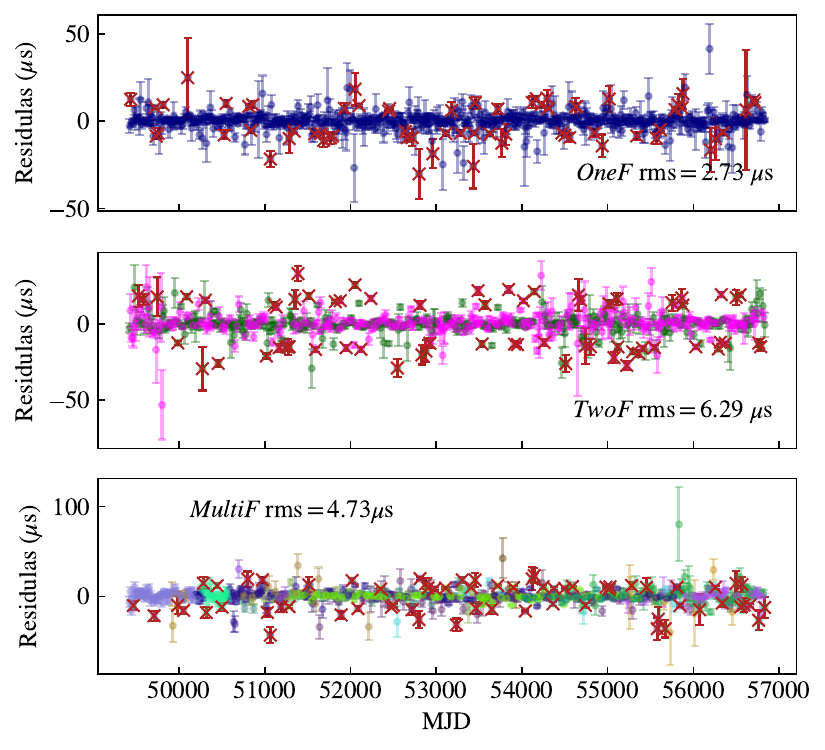}
    \caption{\small{The simulated timing residuals (colored circles)
        of PSR~J1730$-$2304 with $10\%$ of outliers injected (red
        crosses). The different colors of the timing residuals
        represent the systems responsible for the observations. The
        top plot represents the timing residuals simulated for the
        \textit{OneF} dataset, for which we employ a single
        system/observation frequency for all the observations. In the
        central plot we show those for the
        \textit{TwoF} dataset for which we employ two
        systems/observation frequencies, and the bottom plot
        those for the \textit{MultiF}
        dataset for which we consider several systems/observation frequencies. %
        }}
  \label{fig:timig_residuals_example}
\end{figure}

Finally, we add the GWB contribution to complete the datasets. As
shown in \citet{phinney+2001}, the strain spectrum of GWB is expected
to be well modelled by a power-law
\begin{equation}
    h_c(f)=A_{GWB}\left(\frac{f}{1 \mbox{
        yr}^{-1}}\right)^{\alpha_{GWB}}
\end{equation}
where $A_{GWB}$ and $\alpha_{GWB}$ are the GWB strain amplitude and
spectral index, respectively. The corresponding PSD
$\mathcal{P}_{GWB}$ can be parameterized as:
\begin{equation}
   \mathcal{P}_{GWB}(f)=\frac{h_c^2(f)}{12
     \pi^2f^3}=\frac{A_{GWB}^2(f)}{12\pi^2}\left({}\frac{f}{\mbox{
       yr}^{-1}}\right)^{-\gamma_{GWB}}
\end{equation}
with $\gamma_{GWB} = 3-2\alpha_{GWB}$. We set
$A_{GWB}=2\times10^{-15}$, which is consistent with the current PTA
estimates \citep{ng15yr,eptadr2III,pptadr3,cptadr1} and we consider
$\gamma_{GW}=13/3$, which is expected for a GWB generated by a
population of SMBHBs on circular orbits whose evolution is driven by
GW emission \citep{phinney+2001}. For this signal, we fix the number
of Fourier modes to $5$ as done in \cite{Arzoumanian+2020, ng15yr}. After generating the datasets, we corrupted them by injecting outlier
observations. Simulated timing residuals 
follows a Gaussian distribution with zero mean and variance
$\sigma$. We define as outliers a small amount of randomly chosen data
that follows the same distribution but with a very different variance,
$\sigma_{out}$. As shown in \cite{Wang+2021} an outlier indicator
$z_i$ can be used to describe a corrupted dataset:
\begin{equation}
z_i=
    \begin{cases}
      1 & \mbox{outlier } \\ 0 & \mbox{otherwise }
    \end{cases}.
\end{equation}
In this way it is possible to express the $i$-th timing residuals $r_i$
of a pulsar as:
 \begin{equation}\label{y_ioutliers}
    r_i=r_i+z_i\,\sigma_{i,out}.
\end{equation}
Following this definition, we assign $z_i=1$ to a certain percentage
of randomly selected ToAs per pulsar, and we chose the value of
$\sigma_{i,out}$ such that the outliers have no relation with the
majority of the data. Here $\sigma_{i,out}$ is defined as
$\sigma_{i,out}=\alpha \,\sigma_i $ where $\alpha$ is a positive or
negative random number with absolute value $ \in [3,\;5]$ and $\sigma_i$ is the
post-fit timing residual root-mean-square (rms). The percentages of
outliers tested were $0\%$ (i.e., uncorrupted data), $0.3\%$, $1\%$, $5\%$ and $10\%$. In Figure \ref{fig:timig_residuals_example} we show, as an example, the
timing residuals (colored circles) of PSR~J1730$-$2304 for the three
datasets simulated and with $10\%$ outliers injected (red crosses).
\subsection{Datasets for model selection}
\label{sec:modelsel_data}

To conduct the model selection analysis, we employed simulated
datasets that were produced in a manner similar to that of the
datasets given in Section \ref{sec:datasetsSR},\textit{ but without injecting
the GWB}. Similarly, three separate datasets have been produced
(\textit{OneF}, \textit{TwoF}, \textit{MultiF}) and subsequently
tainted with outliers.
\subsection{Statistical inference}
In order to gauge the impact of outliers on the recovery of the signal
injected, we first examine the simulated outliers-corrupted datasets
and we estimate the parameters describing the noises of interest (RN, DM and GWB)
using a PTA-specific Bayesian inference method 
\citep[][]{vanHaasteren+2012,ellis+2017,enterprise} as employed in \citet{Perera+2019,Arzoumanian+2020,Chen+2021,Goncharov+2021b,68pulsar_nanograv,eptadr2III} and
others. Then, we search for,  along with the other signals, an additional common uncorrelated process (CP),
which we modelled as a power-law with an amplitude $A_{CP}$ and a
spectral index $\gamma_{CP}$, considering 30 Fourier modes.
This noise behaves exactly as the pulsar RN, but with the main difference that the amplitude and the spectral index are the same for each pulsar (in the same fashion as the gravitational signal), but without including any spatial correlation. In this way we test  whether the presence of outliers can also
introduce a spurious common process.%

Bayesian inference is based on the Bayes theorem, which states that in
order to obtain the posterior probability distributions of the
parameters of interest, the \textit{likelihood} and the parameters'
prior probability distributions have to be specified.  In terms of
the latter, we kept the WN parameters (EFAC and EQUAD) fixed to the
injected values (see Table \ref{tab:efac_eqaud}), and we used uniform
priors for the RN, DM-induced noise, GWB and CP spectral indices ($\gamma$ $\in$
$[0, 7]$) and log-uniform priors for their amplitudes
($\log \,A_{RN,DM,CP}$ $\in$ $[-20, -10]$, $\log \,A_{GWB}$ $\in$ $[-18, -13]$). The choice of these
distributions closely follows \cite{Arzoumanian+2020} and
\cite{Chen+2021}.

\begin{figure}[h]
  \centering
   \includegraphics[width=\columnwidth]{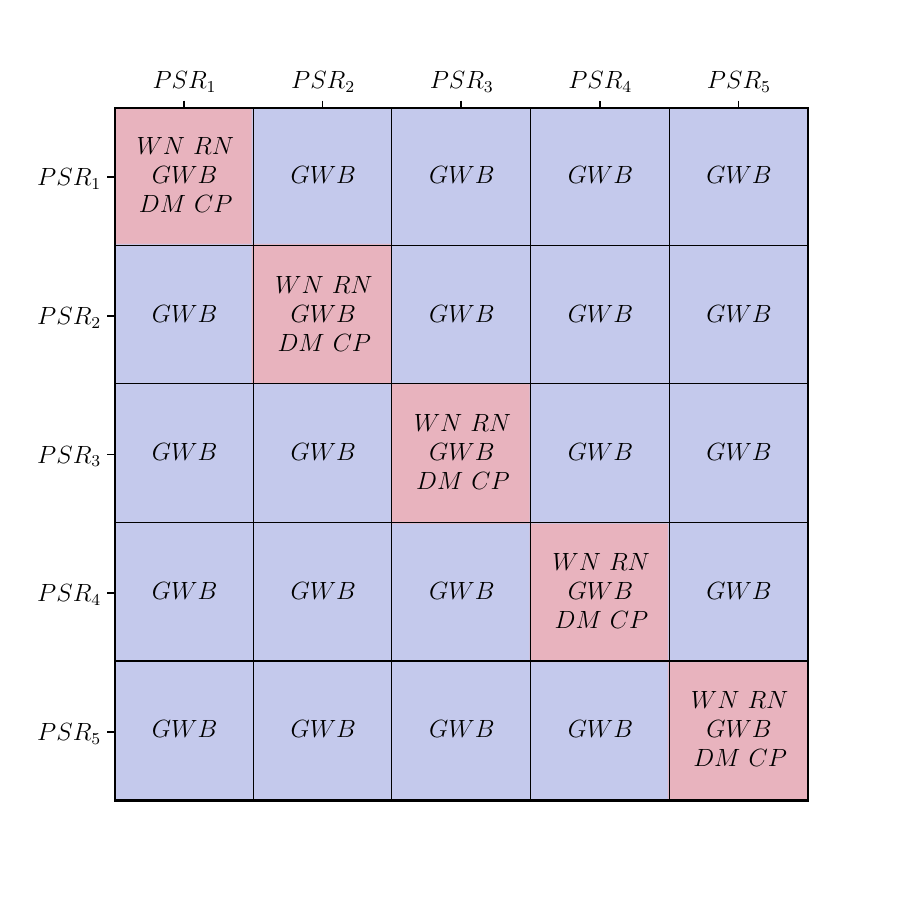}
    \caption{\small{A schematic representation of the covariance
        matrix for five pulsars ($PSR$).   On the diagonal (auto-correlation) the 
        contribution of the WN, RN, DM, CP and GWB signals are present while, in the off-diagonal parts (cross-correlations), there is just that of the GWB.}}
  \label{fig:covmatrixdraw}
\end{figure}

The likelihood can be constructed by assuming that the timing
residuals $r_{ai}$ of the array's $a$-th pulsar, measured at the
$i$-th time, are made up of a deterministic $r_{ai}^{det}$ and a
stochastic component. The former includes, for example, the effects
due to the pulsar spin-down, the annual variations due to the poor
knowledge of the pulsar positions in the sky, the uncertainties in the
location of the Solar System barycentre (SSB), and the phase offsets
or JUMPs due to changes in the equipment, i.e., all the effects that
can be modelled and included in the timing model. The latter includes
the contributions from the intrinsic RN and WN processes,
the DM-induced noise, the clock noise, any common and uncorrelated noise process and the GWB
signal. Therefore, it is possible to write 
\begin{equation}
    r_{ai}=r_{ai}^{det}+r_{ai}^{N}+r_{ai}^{CP}+r_{ai}^{GWB},
\end{equation}
 where $r_{ai}^{CP}$ is the contribution related to an eventual uncorrelated CP, $r_{ai}^{GWB}$ is the stochastic GWB contribution, and $r_{ai}^{N}$ is due to all other
stochastic noise sources. Regarding the latter, we considered only the contributions of the RN, DM-induced noise and WN. Similarly to \cite{maggiore+2008}
\cite{vanHaasteren+2009}, 
\cite{vanHaasteren+2012}, we assumed that both the CP, the GWB and the noise components $N$ are stochastic Gaussian processes, and thus they are fully characterized by their two-point correlation functions that can be
represented by the covariance matrices:
\begin{equation}
    \begin{aligned}
       \langle
        r_{ai}^{N}\,r_{bj}^{N}\rangle&=C^{N}_{(ai)(bj)},\\
         \langle
        r_{ai}^{CP}\,r_{bj}^{CP}\rangle&=C^{CP}_{(ai)(bj)},\\ 
         \langle
        r_{ai}^{GWB}\,r_{bj}^{GWB}\rangle&=C^{GWB}_{(ai)(bj)}.\\ 
    \end{aligned}
\end{equation}
The timing residuals are then distributed as a multidimensional
Gaussian and the likelihood is defined as:
\begin{multline}
P(\{r_{ai}\}|
\bm{\theta})=\mbox{exp}\Biggl(-\frac{1}{2}\sum_{(ai)\,(bj)
}(r_{ai}-r_{ai}^{det})C^{-1}_{(ai)\,(bj)}\\ \times(r_{bj}-r_{bj}^{det})
\Biggr) \frac{1}{\sqrt{\mbox{det}(2\pi C )}},
\end{multline}\label{eq:likelihood}
where $\bm{\theta}$ includes all the parameters characterizing the
timing model, the RN, the DM-induced noise, the WN, the CP and  the GWB;
$C_{(ai)\,(bj)}$ is the total covariance matrix defined as %
\begin{multline}\label{eq:sumofcovariance}
    C_{(ai)\,(bj)}=\delta_{ab}C_{(ai)\,(bj)}^{WN}+\delta_{ab}C_{(ai)\, (bj)}^{RN}+ \delta_{ab}C_{(ai)\, (bj)}^{DM}+\beta_{ab}C^{CP}_{(ai)(bj)}\\+\alpha_{ab}C^{GWB}_{(ai)(bj)},
\end{multline}
where $\delta_{ab}$ is the Kronecker delta, $\beta_{ab}=1$ for any value of $a$ and $b$ (pulsar indices) due to the uncorrelated but common nature of the CP, $\alpha_{ab}=1$ for $a= b$ while, when $a\neq b$, coincide with the HD function multiplied by $3/2$:
\begin{equation}
    \alpha_{ab}=\frac{3}{2} \frac{1-\cos \theta_{ab}}{2} \ln \left( \frac{1-\cos\theta _{ab}}{2} \right)-\frac{1}{4} \frac{1-\cos \theta_{ab}}{2} +\frac{1}{2}
\end{equation} 
where $\theta_{ab}$ is the relative angle between two pulsars.
 It is important to notice that the intrinsic RN and CP differentiate from the GWB since the latter induces an
inter-pulsar correlation between timing residuals. Therefore, because of GWs, the timing residuals of each pulsar are both time (within the pulsar) and spatially correlated (across the array).  A scheme of the structure of this matrix is shown in Figure \ref{fig:covmatrixdraw}.

We used the Enhanced Numerical Toolbox Enabling a Robust PulsaR
Inference SuitE \citep[\protect\texttt{enterprise},][]{enterprise} to
define the prior probability distributions and construct the
likelihood, and then we used the Parallel Tempering Markov Chain Monte
Carlo \citep[\protect\texttt{PTMCMC},][]{ellis+2017} sampling with
$10^6$ iterations to evaluate the posterior probabilities for the
parameters of interest.
\subsection{Model selection analysis}
This analysis closely follows the methods and the models used in
\citet{Zic+2022} and \cite{Arzoumanian+2020}. We used the software
\texttt{enterprise\_extensions} \citep{enterprise+ext} to build the
models and perform the comparison. In this case we only inject WN, RN, DM-induced noise, and outliers (see Sec.~\ref{sec:modelsel_data}). We then analyze the data with the three different models reported in Table \ref{tab:models}.  Also in this case we modelled the CP, present in CP1 and CP2, as a power law described by an amplitude
$A_{CP}$, a spectral index $\gamma_{CP}$ and 30 Fourier
components. For CP1 we use a log-uniform prior on the amplitude
($\log \,A_{CP}$ $\in$ $[-20, -10]$) and a uniform prior on the spectral
index ($\gamma_{CP}$ $\in$ $[0, 7]$). In the case of CP2, we employed
the same prior as in CP1 for the amplitude but we fixed the spectral
index to $13/3$. The model TN (which stands for timing noise) does not include a common process.
\begin{table}
  \centering
  \caption{\small{Models employed for the models selection
      analysis. The CP used in these models is described by a
      power law characterized by an amplitude $A_{CP}$ and a
      spectral index $\gamma_{CP}$.}}
  	\begin{tabular}{cccl} 
            \hline
            Model& WN &RN (DM)&CP \\
            \hline
            TN& \ding{51}&\ding{51}&-\\
            CP1&\ding{51}&\ding{51}&$A_{CP} $; $\gamma_{CP}$\\
            CP2& \ding{51}&\ding{51}& $A_{CP} $; $\gamma_{CP}=13/3$\\
            \hline
  \end{tabular}
  \label{tab:models}
\end{table}
Having defined the models, we used the product-space approach
\citep{Arzoumanian+2020} to pick the one that better describes the
data, between CP1 and TN and between CP2 and TN. This method involves
creating a new variable: the model index, which is then sampled along
with the parameters of the competing models. By evaluating the
proportion of samples in each bin of the model index parameter, we were
then able to evaluate the posterior odds ratio, following the
\textit{hypermodel} method \cite[see e.g., ][]{HHHL16}. 
\section{RESULTS} \label{sec:RESULTS}

\subsection{GWB and pulsar noise recovery in presence of outliers}\label{sec:Signals recovery with outliers}

\begin{figure}
\centering
\includegraphics[width=\linewidth]{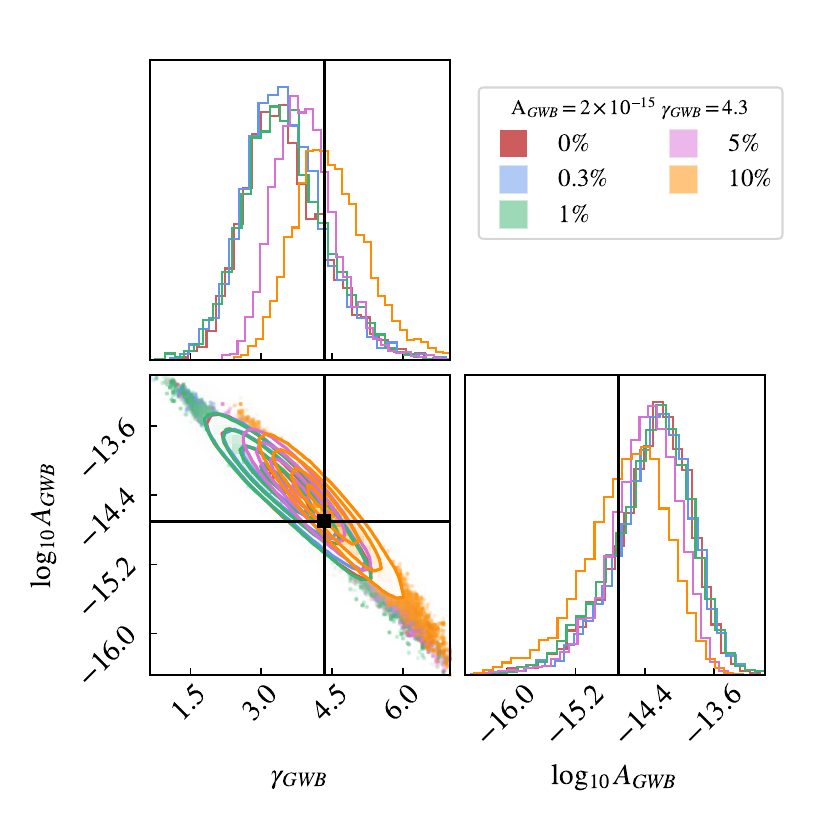}
 \caption{\small{The 2-dimensional marginalized posterior
     distributions of $\log _{10} A_{GWB}$ and $\gamma_{GWB}$ recovered
     for the \textit{OneF} dataset corrupted with $0\%$ (red),  $0.3\%$ (blue), $1\%$ (green), $5\%$ (pink) and $10\%$ (orange) of outliers. Each pair of
     distributions ($\log _{10} A_{GW}$; $\gamma_{GW}$) has been
     recovered separately and then overlapped to be easily
     compared. The black lines and the square indicate the injected
     values of the amplitude and spectral index. %
     }} \label{fig:GWB+OneF+noCP}
\end{figure}

\begin{figure}
 \includegraphics[width=\linewidth]{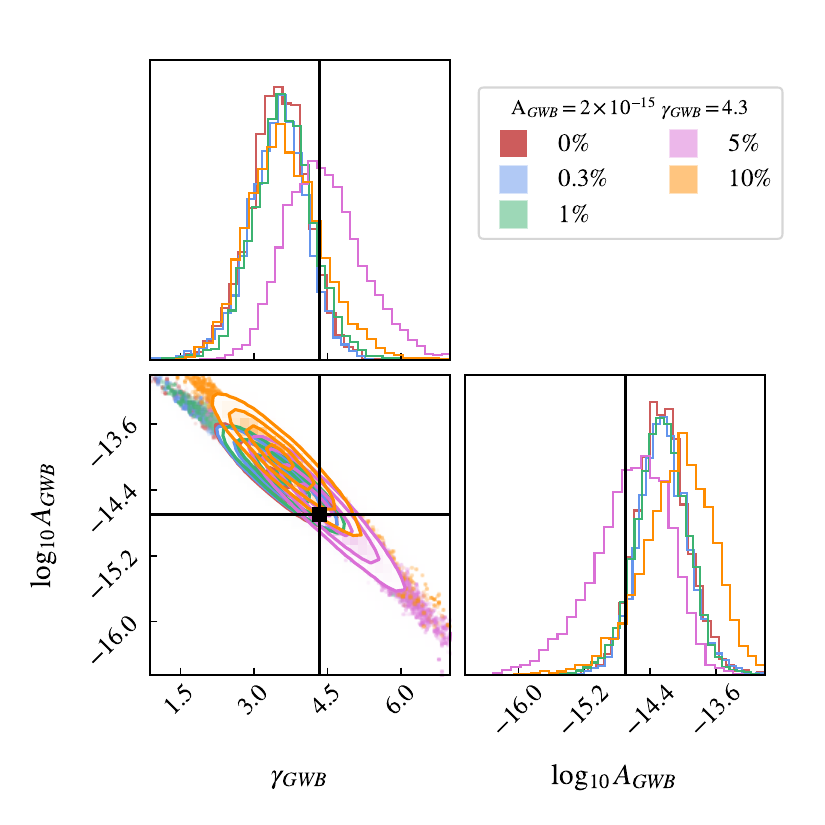}
         \caption{\small{Same as Figure \ref{fig:GWB+OneF+noCP} but
             for the \textit{TwoF} dataset.}}
          \label{fig:GWB+TwoF+noCP}
\end{figure}

\begin{figure}
   \includegraphics[width=\linewidth]{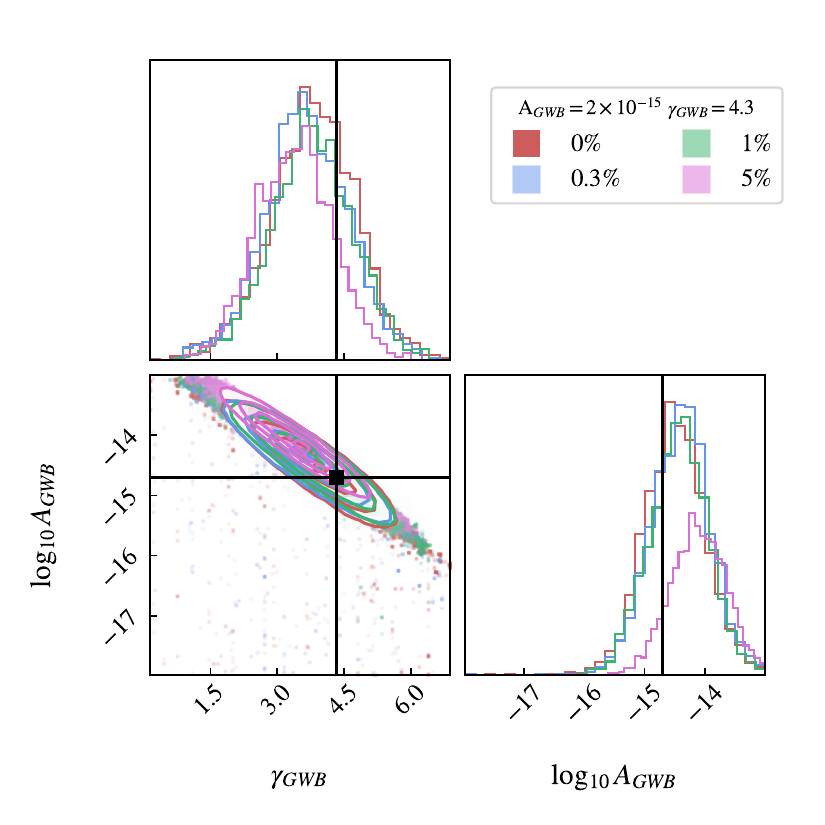}
          \caption{\small{Same as Figure \ref{fig:GWB+OneF+noCP} but
              for the \textit{MultiF} dataset. In this case, the
              results for $10\%$ of outliers injected are missing. Probably due to the high complexity of such dataset we were not able to obtain 
              robust results for this case. However, it is very
              improbable to have data containing such a high percentage of
              outlier and the other percentages analyzed can be
              considered sufficient to study the influence of outliers
              in this dataset.}}
           \label{fig:GWB+MultiF+noCP}
\end{figure}

The results obtained from the runs which consider a model with WN (fixed),
RN, DM-induced noise (for \textit{MultiF} dataset only) and GWB are summarized in Table~\ref{tab:ressum}.
The effect of outliers on the GWB parameter recovery for the three datasets
are reported in Figure \ref{fig:GWB+OneF+noCP}, \ref{fig:GWB+TwoF+noCP} and
\ref{fig:GWB+MultiF+noCP}.
We found the parameters describing the GWB to be, at worst, only
weakly affected by the presence of outliers in any percentage
studied. In fact, it has been possible to recover values of $A_{GWB}$
and $\gamma_{GWB}$ consistently with those injected, within the $95\%$ credible interval.%
The recovery occurs correctly and independently
from the degree of realism of the dataset. Thus, these results show
that the amount %
of outliers injected in these datasets
is not enough to consistently affect the recovery of the parameters
describing the GWB signal. Although some effects can be observed, they
are limited principally to a slight broadening of the posterior distributions or
to a small shift away from the expected center and they are observed only
when $5\%$ and $10\%$ outliers are injected. The results of the analysis
of the data with $10\%$ injected outliers are not shown in Figure
\ref{fig:GWB+MultiF+noCP}. With such a large fraction of outliers,
the examination of the \textit{MultiF} chains revealed sampling
issues, making it difficult to produce reliable results. We believe 
the other percentages studied to be adequate to analyze the impact of
outliers in the \textit{MultiF} dataset since it is highly
improbable that such a percentage ($10\%$) of outliers could be present
in real PTA datasets.

\begin{figure*}
\includegraphics[width=\textwidth]{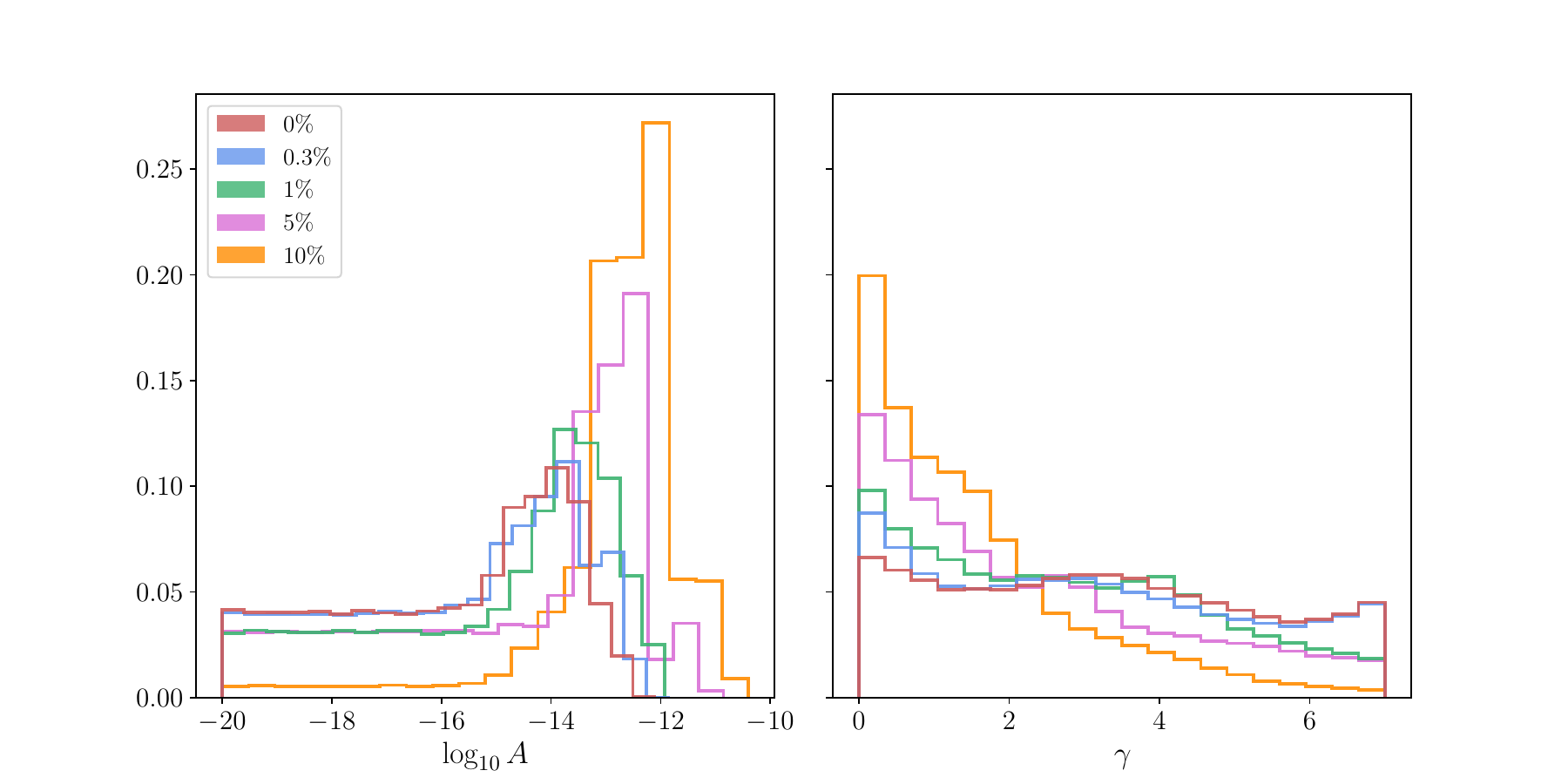}
\caption{\small{The cumulative marginalized posterior distributions of the amplitudes (left) and the spectral
  indices (right) of pulsars' RNs  for the dataset
  \textit{OneF} corrupted with $0\%$ (red),  $0.3\%$ (blue), $1\%$ (green), $5\%$ (pink) and $10\%$ (orange) of outliers. The histograms, in both panels, are  the sum of the normalized marginalized posteriors of the amplitudes and the spectral indices of the RN of each pulsar.}}
  .%
\label{fig:A_G_evoul_onef}
\end{figure*}

\begin{figure*}[h]
\centering
\includegraphics[width=\textwidth]{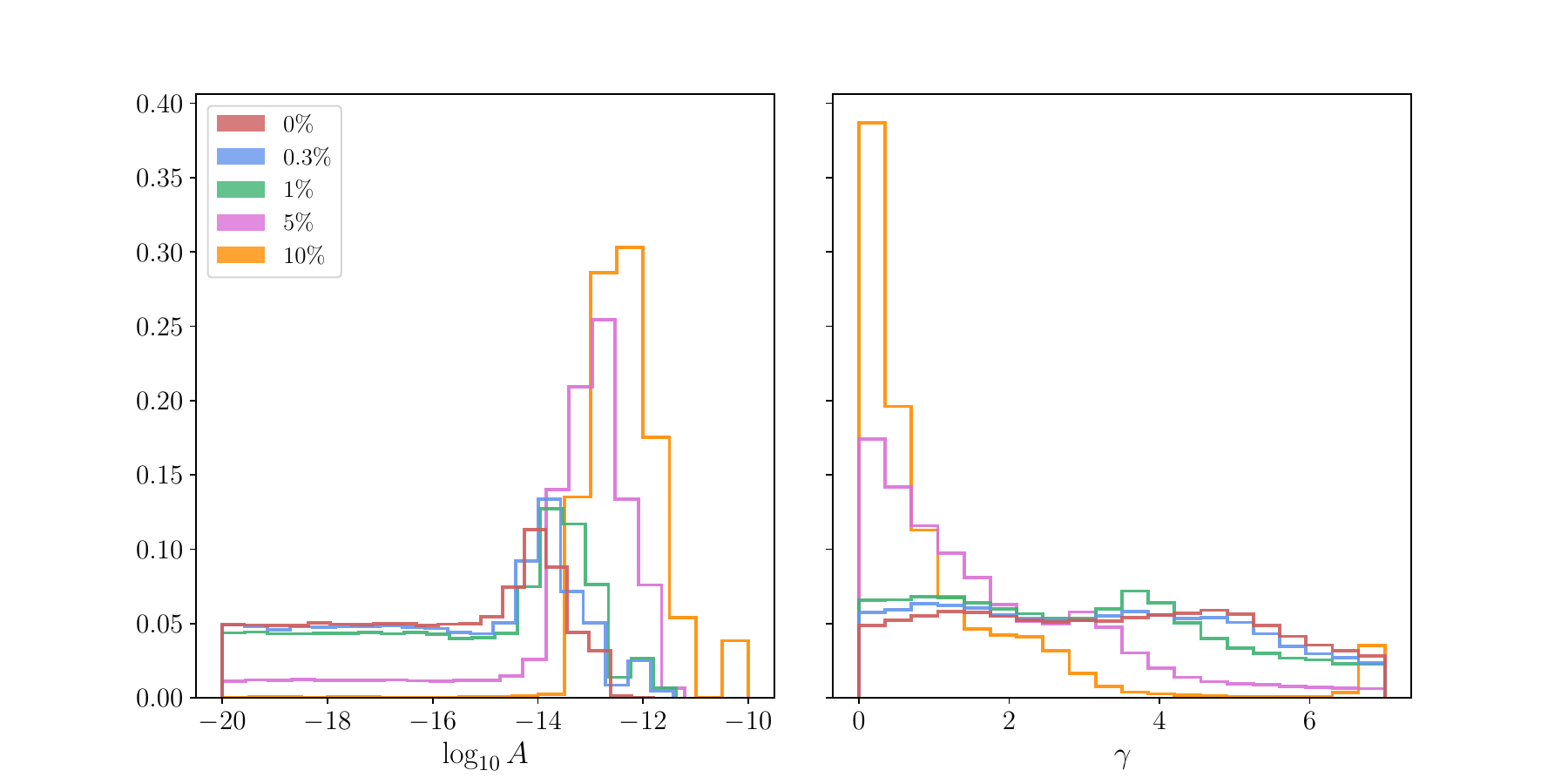}
\caption{\small{Same as Figure \ref{fig:A_G_evoul_onef} but for the dataset
  \textit{TwoF}.}}
  \label{fig:A_G_evoul_twof}
\end{figure*}

\begin{figure*}[h!]
\centering
\includegraphics[width=\textwidth]{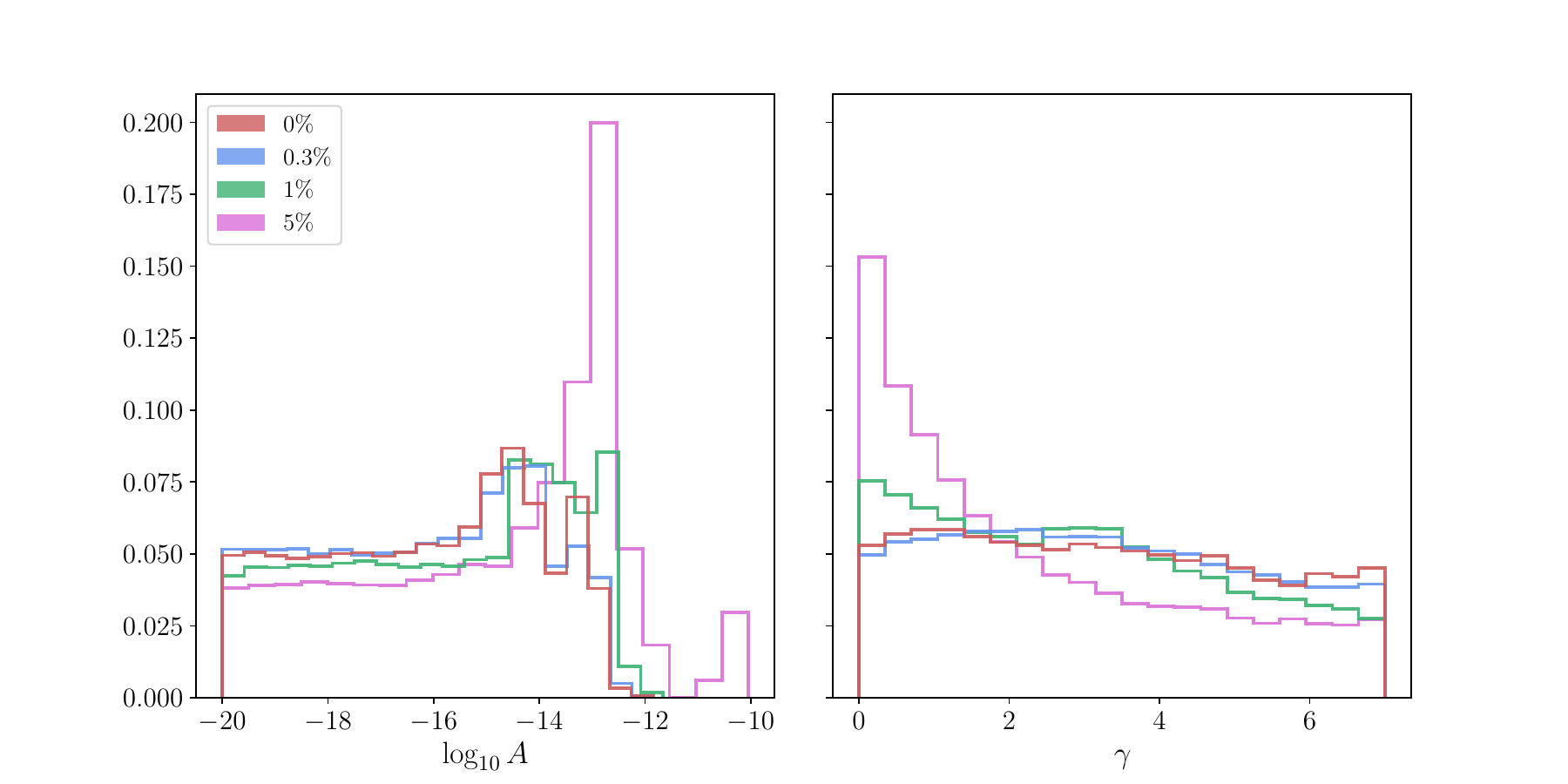}
\caption{\small{ Same as Figure \ref{fig:A_G_evoul_onef} but for the dataset
  \textit{MultiF}. As done in Figure~\ref{fig:GWB+MultiF+noCP}, the distributions for the data corrupted with $10\%$ of outliers are not reported. }}
\label{fig:A_G_evoul_multif}
\end{figure*}

In contrast to the GWB, the recovery of the parameters describing the
pulsars RNs  and DM-induced noises is strongly affected by outliers.
 Regarding the RNs, already with $1\%$ of outliers, the recovered posteriors of the amplitudes and of the spectral indices systematically shift with respect to those recovered from the uncorrupted datasets. In particular those of the amplitudes tend to move toward the
upper limit of the prior range employed in the analysis, while those of the spectral
indices tend to move toward the lower limit. For both these parameters, the
posteriors tend to became narrower as the number of outliers increases.
In Figures \ref{fig:A_G_evoul_onef}, \ref{fig:A_G_evoul_twof}, and \ref{fig:A_G_evoul_multif}, we present cumulative marginalized posterior distributions for the amplitudes and spectral indices of the RNs across the three datasets under consideration. These distributions illustrate the cumulative effect of varying percentages of outliers.  Specifically, each histogram, corresponding to a certain outlier percentage, represents the sum of all normalized marginalized posteriors of the amplitude or spectral index of the pulsars' RNs.
As can be observed, as the percentage of outliers increases the cumulative distributions of $\log_{10} A$ and $\gamma$ move, respectively, toward higher and lower values, becoming narrower and narrower. This imply that, in the presence of outliers, the intrinsic RN of pulsars is recovered 
almost as a higher-amplitude-WN since the spectral indices generally tend to cluster around $0$ and the amplitudes tend to increase of almost 2 orders of magnitude. 

\begin{figure*}[h!]
\centering
\includegraphics[width=\textwidth]{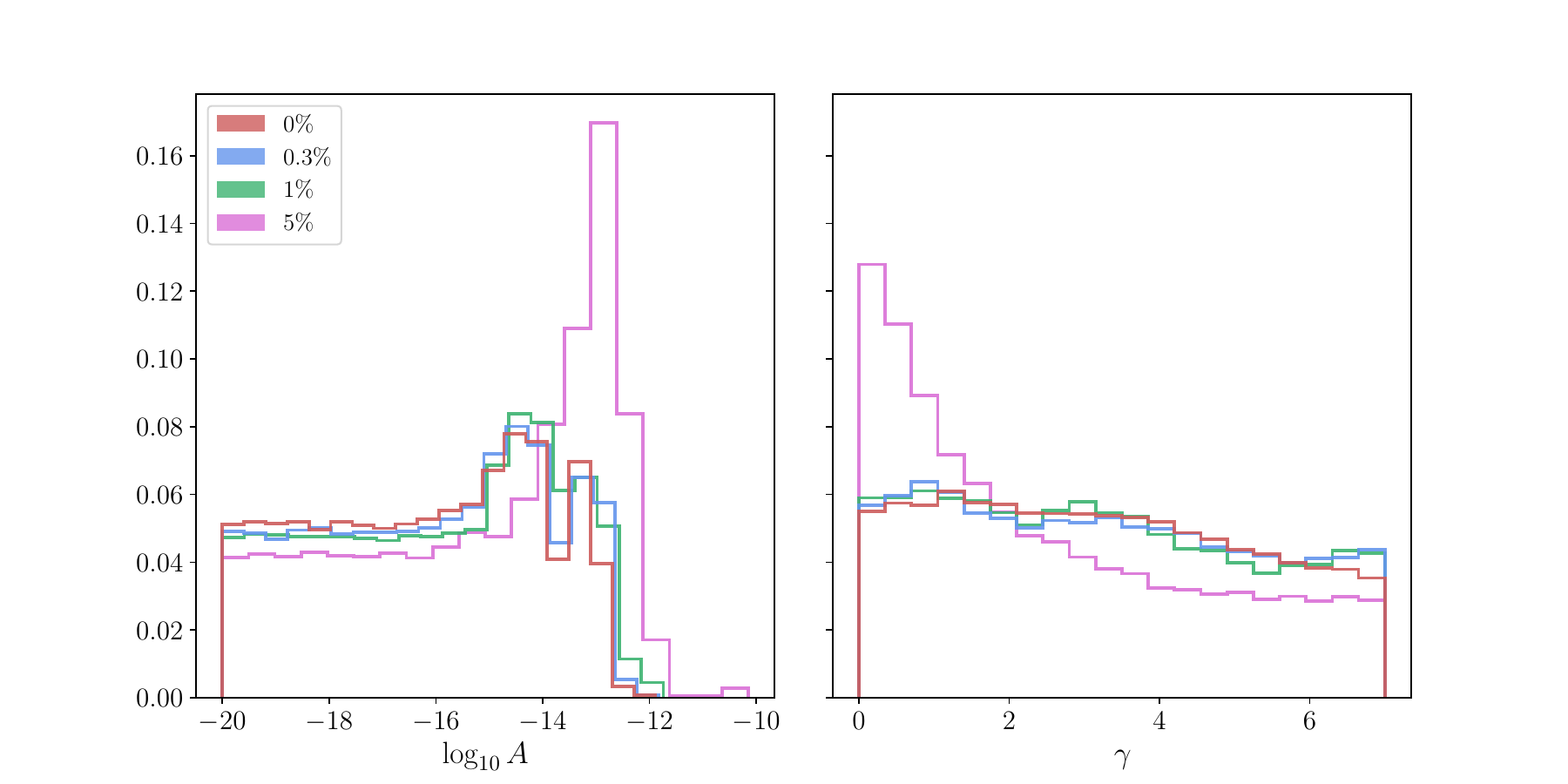}
\caption{ \small{The cumulative marginalized posterior distributions of the amplitudes (left) and the spectral
  indices (right) of the DM-induced noise specific for each pulsar for the dataset \textit{MultiF} corrupted with $0\%$ (red),  $0.3\%$ (blue), $1\%$ (green), $5\%$ (pink) of outliers. The histograms, in each panel, are  the sum of the normalized marginalized posteriors of the amplitudes and the spectral indices of the DM-induced noises of the pulsars. As done in Figures~\ref{fig:GWB+MultiF+noCP},~\ref{fig:A_G_evoul_multif}, the distributions for the data corrupted with $10\%$ of outliers are not reported.}}
\label{fig:A_G_DM_evoul_multi}
\end{figure*}

We observe a similar trend, albeit less pronounced, for the recovered DM-induced noise in the \textit{MultiF} dataset, as depicted in Figure \ref{fig:A_G_DM_evoul_multi}. %
In the case of uncorrupted data, the amplitudes and spectral indices are weakly constrained, and the shift towards higher amplitudes and smaller spectral indices, due to outliers, is less prominent.
This can be attributed to the challenging nature of recovering this signal, primarily due to its frequency-dependent characteristics. Successful constraining would require multiple observations at various frequencies for each epoch. The \textit{MultiF} dataset is designed to emulate real EPTA data, where achieving an optimally diverse frequency coverage is often unfeasible. This inherent lack of sensitivity across the entire observed time span constrains our ability to accurately recover the DM models. 

\subsection{Spurious common process due to outliers}\label{sec:Common process recovery}
\begin{figure*}[h]
\centering
\begin{minipage}[b]{.49\textwidth}
\includegraphics[scale=0.65]{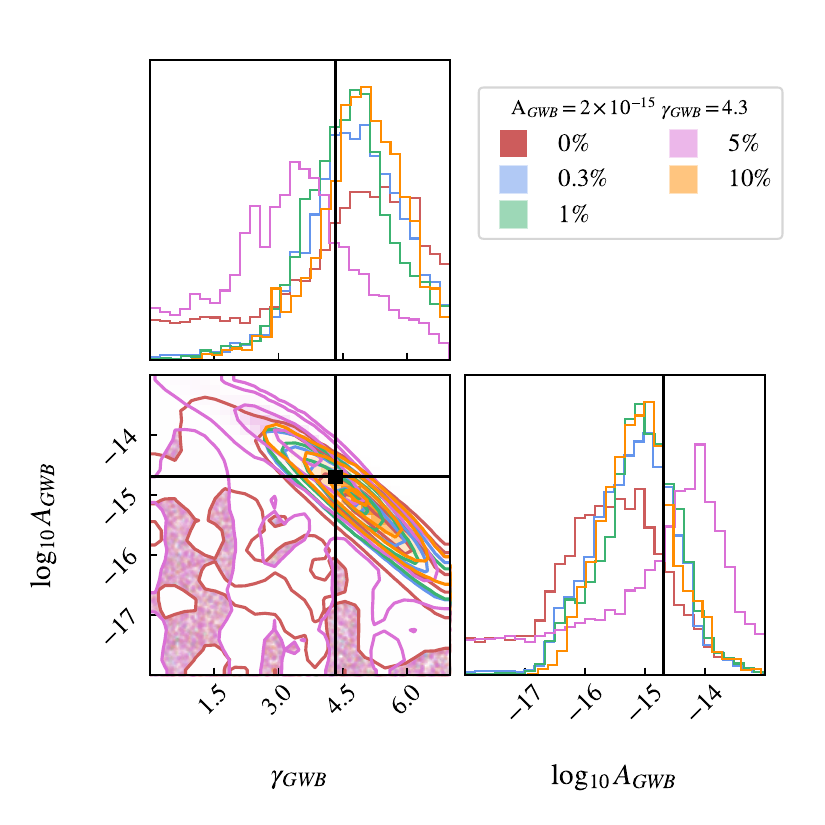}
\end{minipage}\hfill
\begin{minipage}[b]{.49\textwidth}
\includegraphics[scale=0.65]{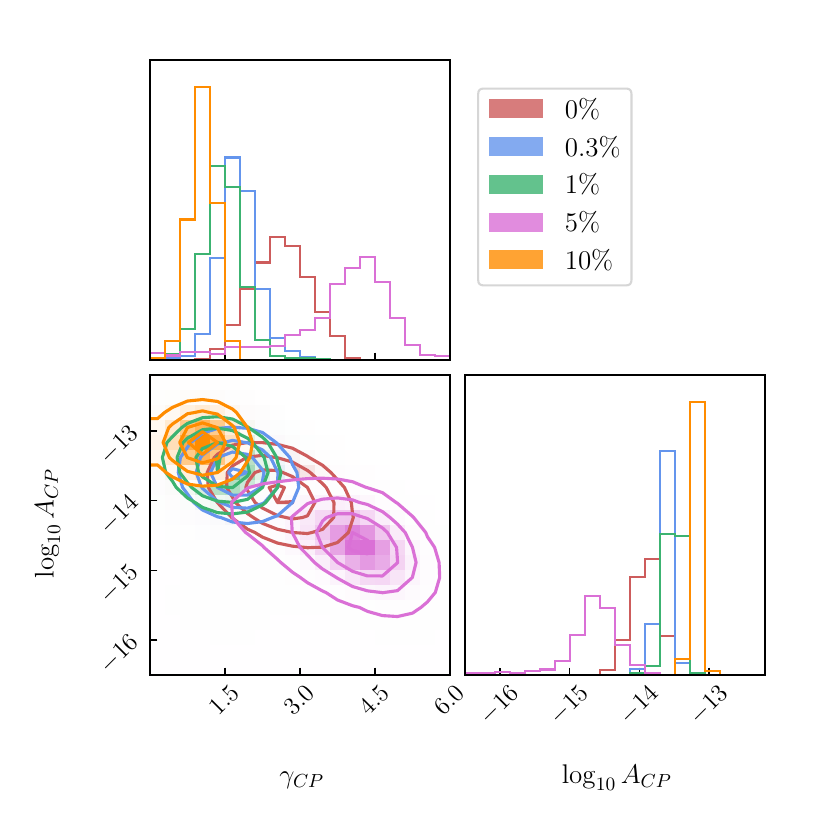}
\end{minipage}
\caption{\small{ The recovered common processes searched in the \textit{OneF}
    dataset corrupted with $0\%$ (red),  $0.3\%$ (blue), $1\%$ (green), $5\%$ (pink) and $10\%$ (orange) of outliers. \textit{Left}: the two-dimensional posterior probabilities
    of the parameters characterizing the GWB ($ A_{GWB}$,$\gamma_{GWB}$). The injected values ($2 \times 10^{-15}$, $4.3$) are represented by black lines and a square symbol. Notably, in this case, the signal can consistently be accurately recovered. \textit{Right}: the two-dimensional posterior probabilities
    of the parameters characterizing the CP ($ A_{CP}$,$\gamma_{CP}$). Unlike the GWB, this signal was not directly injected into the data.}}
\label{fig:CP+GWB_onef}   
\end{figure*}

\begin{figure*}
\centering
\begin{minipage}[b]{.49\textwidth}
\includegraphics[scale=0.65]{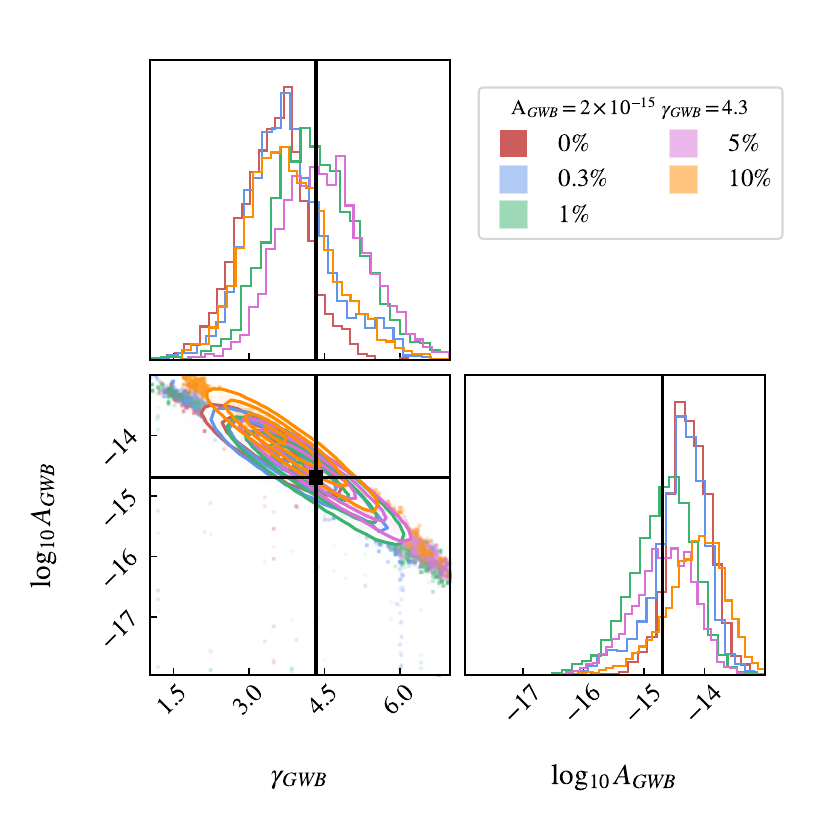}
\end{minipage}\hfill
\begin{minipage}[b]{.49\textwidth}
\includegraphics[scale=0.65]{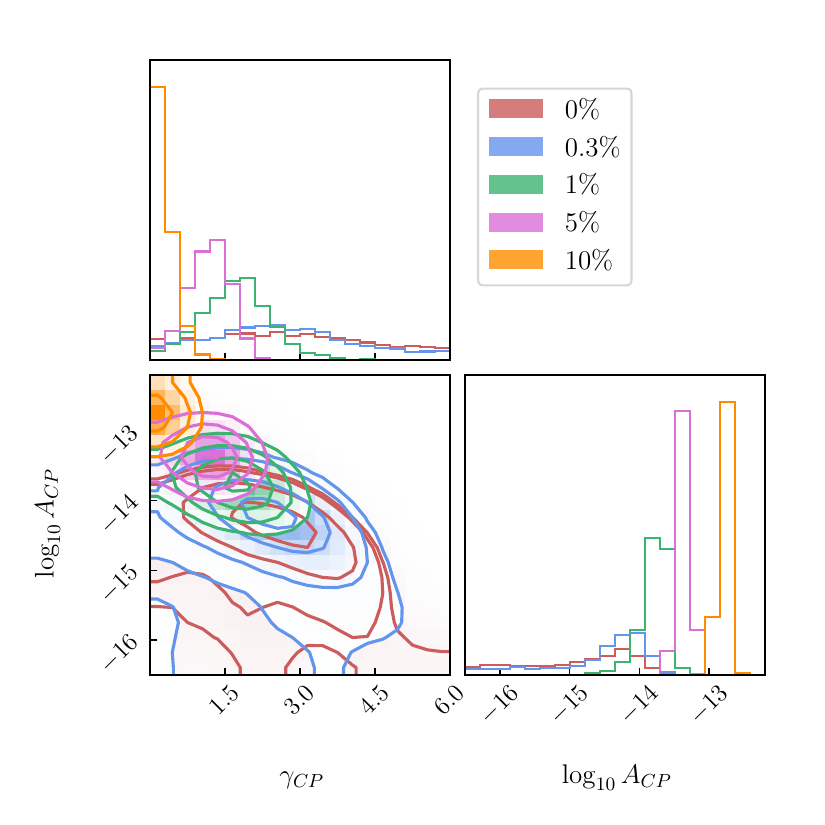}
\end{minipage}
\caption{\small{Same as Figure \ref{fig:CP+GWB_onef} but for the
    \textit{TwoF} dataset.}}
\label{fig:CP+GWB_twof}
\end{figure*}

\begin{figure*}
\centering
\begin{minipage}[b]{.49\textwidth}
\includegraphics[scale=0.65]{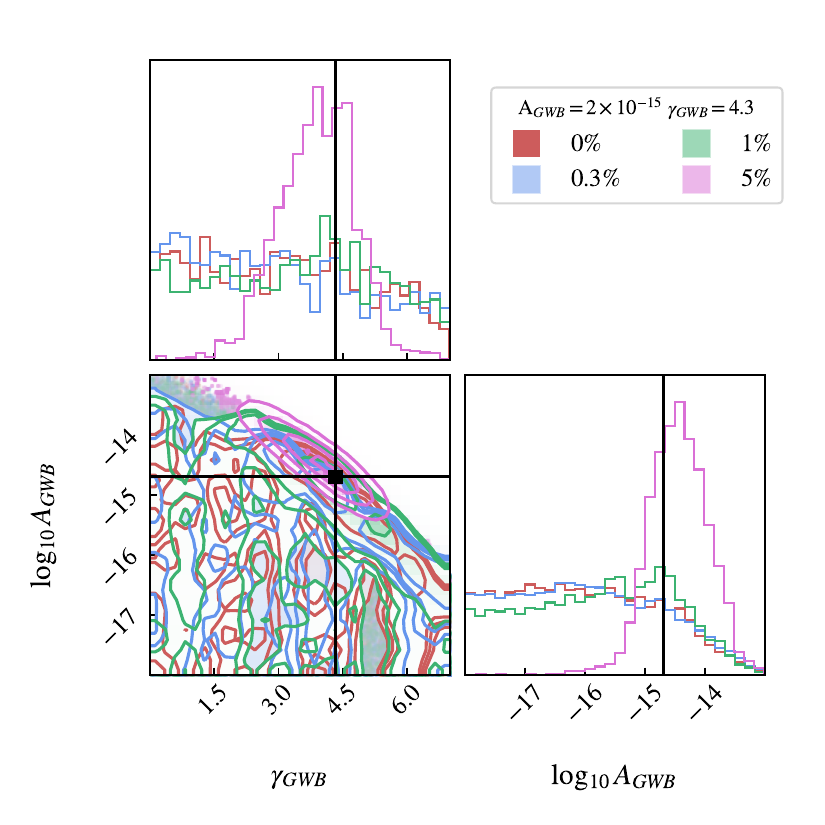}
\end{minipage}\hfill
\begin{minipage}[b]{.49\textwidth}
\includegraphics[scale=0.65]{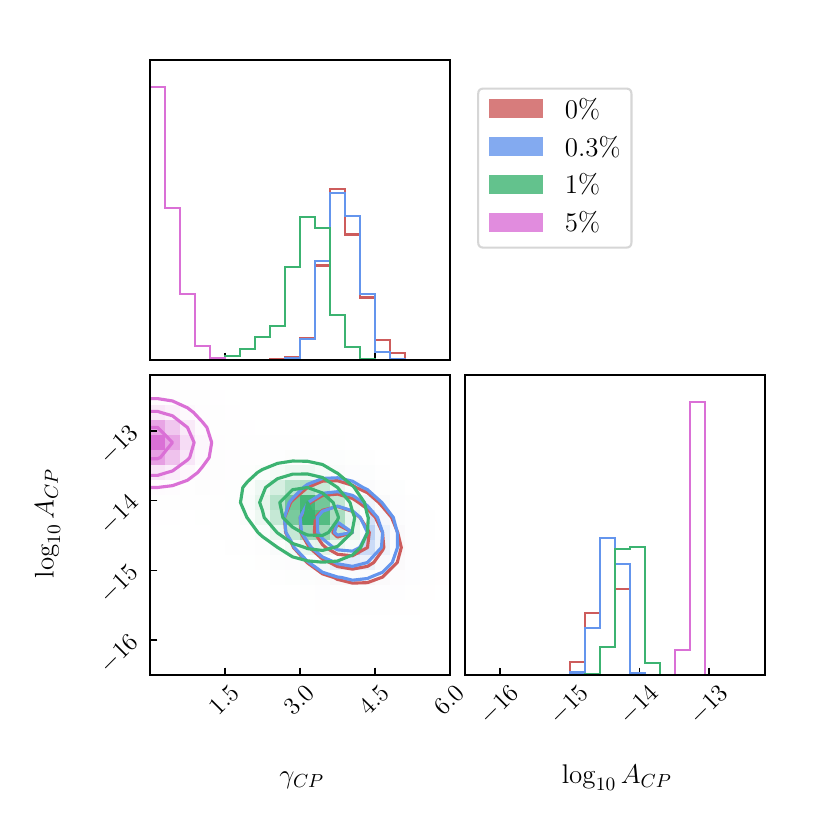}
\end{minipage}
\caption{\small{Same as Figure \ref{fig:CP+GWB_onef} but for the
    \textit{MultiF} dataset. In this scenario, when searching for the GWB in conjunction with an uncorrelated CP, successful recovery is not achievable when the data contains less than $5\%$ outliers. However, with the presence of $5\%$ outliers, successful recovery becomes possible.}}
\label{fig:CP+GWB_multif}
\end{figure*}

Once we established the effects of outliers on the recovery of an injected GWB and intrinsic pulsar noises, we checked whether the presence of outliers can lead to the spurious detection of an uncorrelated CP. To this aim, we considered the same data used in Sec.~\ref{sec:Signals recovery with outliers} (which include a GWB and outliers) but we added an uncorrelated CP, modelled as a power law ($A_{CP}$; $\gamma_{CP}$), to the recovery model. We also added, for each dataset (\textit{OneF}, \textit{TwoF}, \textit{MuliF}), a test run in which we consider data with no outliers and no GWB injected and perform a search for RN, DM and CP by fixing the WN parameters. This was done to check whether a CP could emerge in  datasets that are not corrupted by outliers and in which no correlated common signal (e.g. a GWB) is present.%
\begin{table}
  \centering
  \caption{\small{Summary of the recovery performance of the
      uncorrelated CP and the GWB for the three datasets studied.
      See Section~\protect\ref{sec:Common process recovery} for details.}\label{tab:ressum}}
  \begin{tabular}{@{}c*{8}{c}@{}}
    \hline
    Outlier& Dataset & \multicolumn{1}{c}{GWB} & \multicolumn{1}{c}{CP} &
    \multicolumn{2}{c}{GWB+CP} & \multicolumn{1}{c}{DM} & \multicolumn{1}{c}{RN} \\
    \cline{5-6}
       \%  & & & & GWB & CP & & \\
    \hline
    & 1F&       \ding{51}&\ding{55}&\ding{51}&\ding{51}& \ding{51}&\ding{51}&\\
 0.0& 2F&       \ding{51}&\ding{55}&\ding{51}&\ding{51}& \ding{51}&\ding{51}&\\
    & MF&       \ding{51}&\ding{55}&\ding{55}&\ding{51}& \ding{51}&\ding{51}&\\
    & MF+10yr&  \ding{51}&\ding{55}&\ding{51}&\ding{51}& \ding{51}&\ding{51}&\\
\thinspace                                                                                 
    & 1F&       \ding{51}&\ding{55}&\ding{51}&\ding{51}& \ding{51}&\ding{51}&\\
 0.3& 2F&       \ding{51}&\ding{55}&\ding{51}&\ding{51}& \ding{51}&\ding{51}&\\
    & MF&       \ding{51}&\ding{55}&\ding{55}&\ding{51}& \ding{51}&\ding{51}&\\
    & MF+10yr&  \ding{51}&\ding{55}&\ding{51}&\ding{51}& \ding{51}&\ding{51}&\\
\thinspace                                                                                 
    & 1F&       \ding{51}&\ding{51}&\ding{51}&\ding{51}& \ding{51}&\ding{51}&\\
 1.0& 2F&       \ding{51}&\ding{51}&\ding{51}&\ding{51}& \ding{51}&\ding{51}&\\
    & MF&       \ding{51}&\ding{51}&\ding{55}&\ding{51}& \ding{51}&\ding{51}&\\
    & MF+10yr&  \ding{51}&\ding{51}&\ding{51}&\ding{51}& \ding{51}&\ding{51}&\\
\thinspace                                                                                
    & 1F&       \ding{51}&\ding{51}&\ding{51}&\ding{51}& \ding{51}&\ding{51}&\\
 5.0& 2F&       \ding{51}&\ding{51}&\ding{51}&\ding{51}& \ding{51}&\ding{51}&\\
    & MF&       \ding{51}&\ding{51}&\ding{55}&\ding{51}& \ding{51}&\ding{51}&\\
    & MF+10yr&  \ding{51}&\ding{51}&\ding{51}&\ding{51}& \ding{51}&\ding{51}&\\
\thinspace                                                                                
    & 1F&  \ding{51}&\ding{51}&\ding{51}&\ding{51}& \ding{51}&\ding{51}&\\
10.0& 2F&  \ding{51}&\ding{51}&\ding{51}&\ding{51}& \ding{51}&\ding{51}&\\
    & MF&  \ding{51}&\ding{51}&\ding{55}&\ding{51}& \ding{51}&\ding{51}&\\
    & MF+10yr&  \ding{51}&\ding{51}&\ding{51}&\ding{51}& \ding{51}&\ding{51}&\\
    \hline
  \end{tabular}
\end{table}
The uncorrelated CPs and the GWB recoveries are reported, for each dataset, in
Figure \ref{fig:CP+GWB_onef}, \ref{fig:CP+GWB_twof} and
\ref{fig:CP+GWB_multif}.
For the \textit{OneF} dataset, %
the GWB can be recovered within the $95\%$ credible interval consistently with the results reported for this dataset in Section \ref{sec:Signals recovery with outliers}.
Alongside with the GWB, it is possible to  recover, independently from the number of outliers, a well-constrained CP which evolution depends on the severity of the contamination -- as the number of outliers increase, the CP recovered moves
toward higher amplitudes and lower spectral indices. This kind of
evolution is the same that has been observed for the RNs and DMs as
reported in Section \ref{sec:Signals recovery with outliers}. 
In contrast to the other datasets, the test-search conducted on the \textit{TwoF} dataset revealed the presence of a CP. This suggests that sources other than the outliers and the GWB might be capable of inducing a CP in this dataset. However, after injecting the GWB, this particular feature is less evident. A CP is again distinctly detected when at least $1\%$ of outliers are injected into the data. Once recovered, this signal follows the same evolution as observed for the \textit{OneF} dataset.
In the case of the \textit{MultiF} dataset, while the measured uncorrelated CP follows the same trends seen in the other datasets, a markedly different behavior can be observed for the GWB signal. When conducting a joint search for the GWB and a CP within this dataset, we observed that well-constrained posterior probabilities for the parameters characterizing the GWB can not be obtained, whereas the opposite holds true for the CP. This phenomenon is particularly prominent when fewer than $5\%$ outliers are introduced into the dataset.
Interestingly, with a $5\%$ outlier presence in the data, it becomes feasible to effectively constrain the GWB.
The latter result may be attributed to the fact that, when a relatively high percentage of outliers is introduced, the CP signal induced becomes distinctly discernible from the GWB signal, allowing the latter to emerge more clearly.
On the other hand, the inability to recover the GWB in the presence of other outlier percentages can be attributed both to the degree of realism of this dataset and to the resemblance between the gravitational signal and the CP.  The GWB signal, as described in Section
\ref{sec:INTRODUCTION}, is formed by an uncorrelated part
(auto-correlation terms along the diagonal of the matrix in Figure
\ref{fig:covmatrixdraw}) and by a correlated part (cross-correlation
terms in the off-diagonal part of the matrix in Figure~\ref{fig:covmatrixdraw}) which is expected to be weaker
with respect to the former, due to the magnitude of the correlation
coefficients ($\alpha_{ab} \leq0.5$ for $a\neq b$). Therefore, it has
been hypothesized that the auto-correlated component of the GWB signal
should be the first to be observed \citep{Romano+2021,Pol+21}.
This, in fact, was confirmed by real data, where a common red signal was first detected, and then evidence for correlation started to emerge.
\citet{Pol+21} demonstrate that effective evidence for the
cross-correlation component could be observed when, the datasets that
they had examined, had a time span of $\sim 18-20$ years. When an uncorrelated CP is searched alongside the GWB in our most realistic
dataset, \textit{MultiF}, it is possible that part of the auto-correlation
component of the GWB flows into the power of the uncorrelated CP and
the cross-correlated part is unable to emerge with sufficient strength to be constrained, resulting in the recovery shown in the left plot of
Figure \ref{fig:CP+GWB_multif}. 
Following similar reasoning as \citet{Romano+21} and \citet{Pol+21} we therefore introduced the dataset \textit{MultiF+10Y}, which consists of the %
\textit{MultiF} dataset with the time span extended by $10$ years. We retained the same number of observations, uncertainties, frequencies, systems, and observatories. In confirmation of our hypothesis, with this dataset it is possible to recover the GWB
within the  $95 \%$ credible interval.
\begin{figure*}
\centering
\begin{minipage}[b]{.49\textwidth}
\includegraphics[scale=0.65]{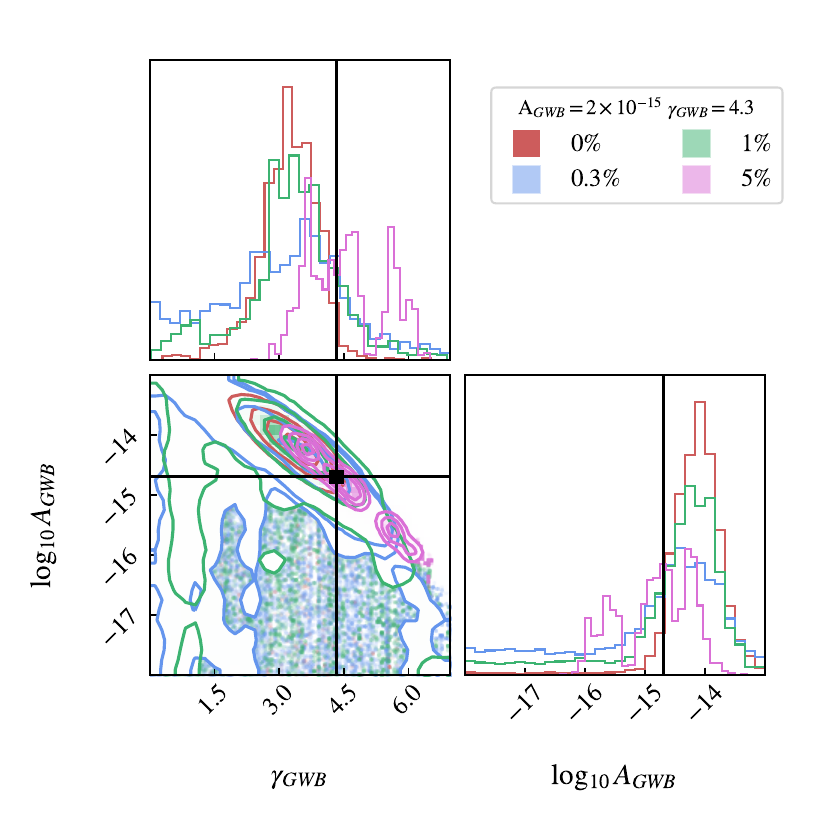}
\end{minipage}\hfill
\begin{minipage}[b]{.49\textwidth}
\includegraphics[scale=0.65]{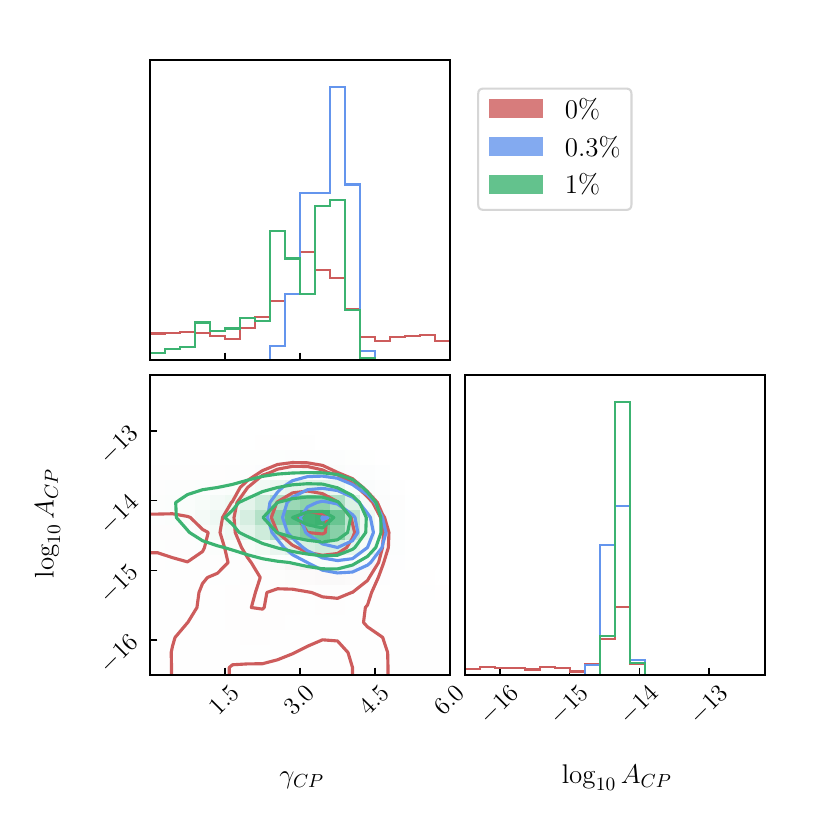}
\end{minipage}
\caption{\small{Same as Figure \ref{fig:CP+GWB_onef} but for the
    \textit{MultiF+10Y} dataset. This dataset is identical to the \textit{MultiF} dataset, with the exception of a time span extended by $10 $ years. This extension significantly enhances sensitivity to the GWB, leading to a more pronounced emergence of the correlated component of the signal (left).
    However, it was not feasible to accurately generate posterior probability distributions for the CP when $5\%$ of outliers were present; hence, these results have not been included (right).}}
\label{fig:CP+GWB_multif+10y}
\end{figure*}
\subsection{Model selection}\label{sec:Model selection}
To further study outliers as possible source of spurius
uncorrelated CP able to contribute to the signal detected in 
real PTA data analysis, we conducted a models comparison-analysis, considering the models reported in Table \ref{tab:models} and employing the datasets presented in Section \ref{sec:modelsel_data}. %
\subsubsection{CP1 vs TN}
\begin{figure*}
\centering
\begin{minipage}[b]{.33\textwidth}\includegraphics[scale=0.43]{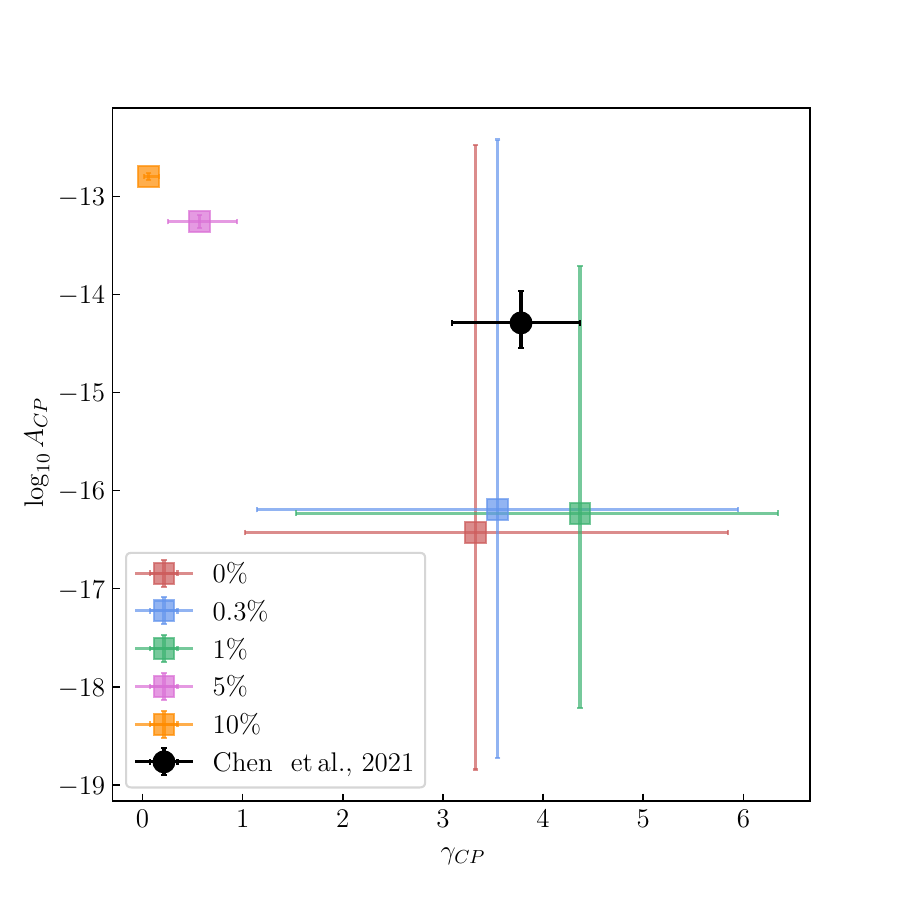}
\end{minipage}
\begin{minipage}[b]{.33\textwidth}
\includegraphics[scale=0.43]{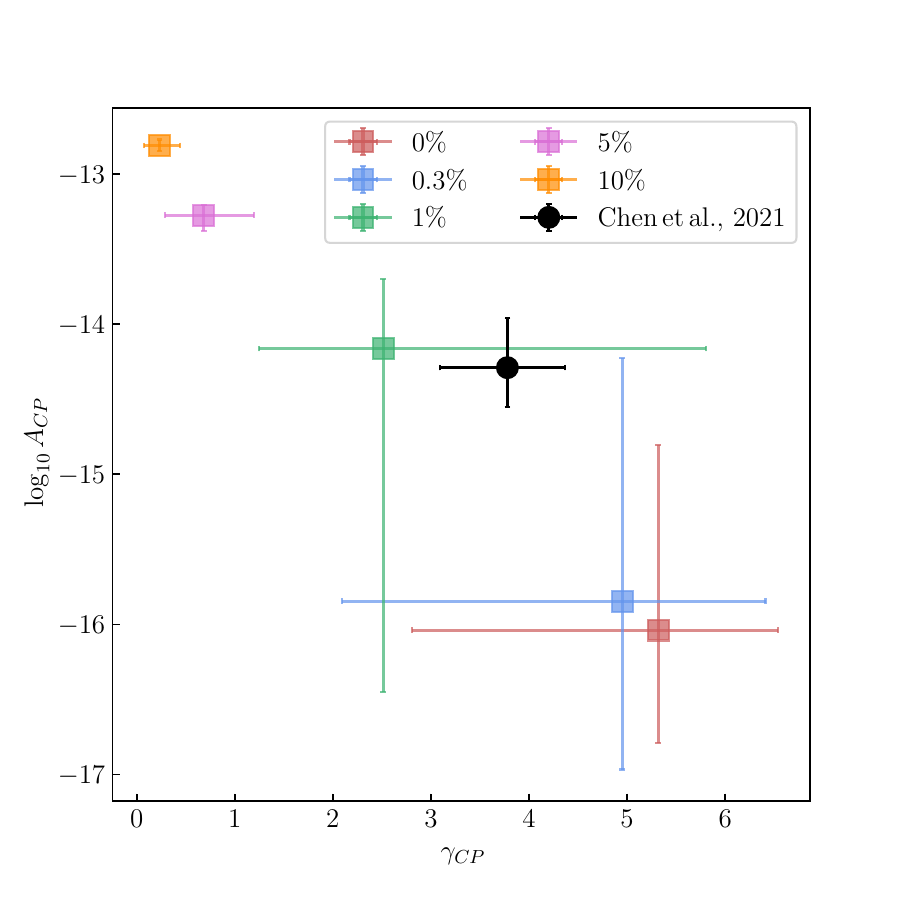}
\end{minipage}
\begin{minipage}[b]{.33\textwidth}
\includegraphics[scale=0.43]{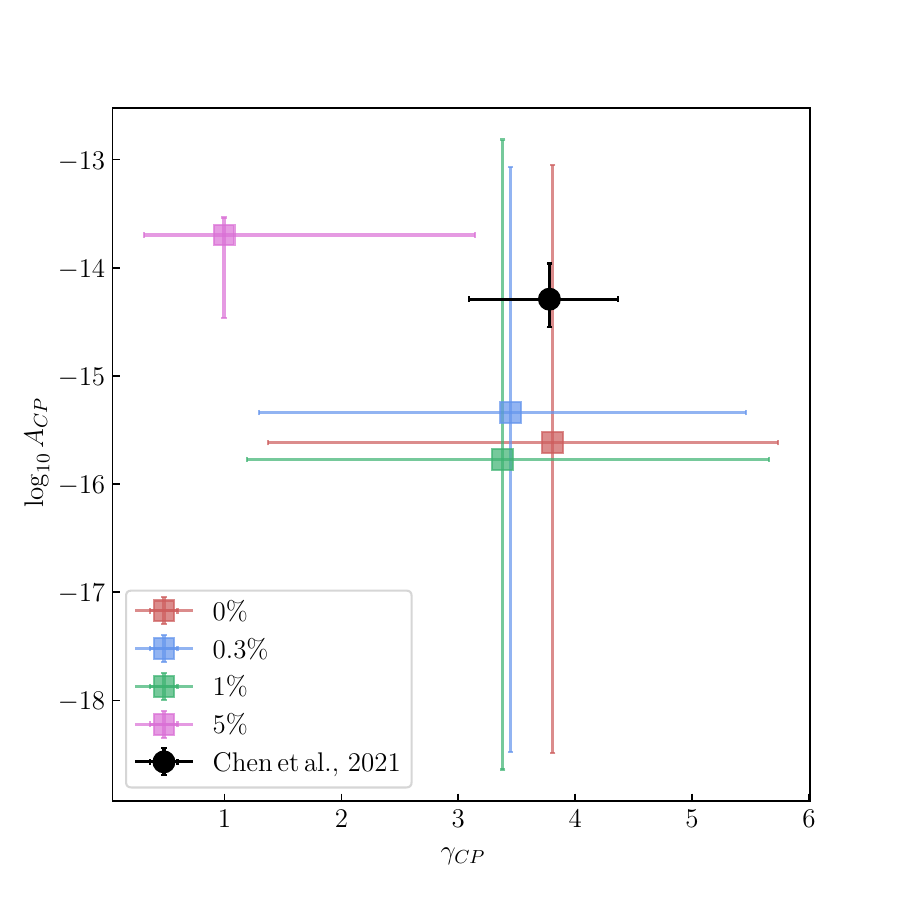}
\end{minipage}
\caption{\small{The uncorrelated CP recovered from the comparison
      between the models CP1 and TN for the datasets 
      \textit{OneF} (left), \ textit{TwoF} (center) and \textit{MultiF} (right) dataset
       corrupted with $0\%$ (red),  $0.3\%$ (blue), $1\%$ (green), $5\%$ (pink) and $10\%$ (orange) of outliers. The error bars represent the 68\% credible interval. The uncorrelated CP recovered, modelling the CP as done in CP1, in \citep{Chen+2021} is represented in black. 
      }}
\label{fig:CP1vsTN}
\end{figure*}

For each dataset considered, we observed increasing evidence in
support of CP1 over TN, with the growth being correlated with the percentage of outliers injected. According to Table \ref{tab:bf_CP1}, in which we have reported the $\log_{10}$ posterior odds ratios resulting from the model comparison, when $0$ to $1\%$ of outliers are injected, there is weak
but gradually growing support for CP1, while when $5\%$ and $10\%$ of outliers are present in the data, this support becomes fairly substantial.
\begin{table}
  \centering
  \caption{\small{$\log_{10}$ posterior odds ratios obtained from the model
      comparison between CP1 and TN for the datasets,
      and percentages of outliers studied. For the
      \textit{MultiF} %
      dataset, this analysis has not been
      performed in the scenario of $10\%$ of
      outliers injected in the data. Uncertainty over the final digit is indicated by the number in parentheses.}}
  \resizebox{\columnwidth}{!}{%
    \begin{tabular}{c*{6}{>{$}c<{$}}}
  \hline
  Dataset& 0\%&0.3\%&1\%&5\%&10\%\\
  \hline
  \textit{OneF}&-0.295(4)&-0.231(5)&0.141(6)&4.9&4.9\\
  \textit{TwoF}&0.598(9)&0.415(8)&0.81(1)&2.9(2)&4.9\\
  \textit{MultiF}&-0.035(8)&-0.057(8)&-0.155(8)&1.67(3)&-\\
  \hline
  \end{tabular}
  }
  \label{tab:bf_CP1}
\end{table}
In Figure \ref{fig:CP1vsTN} are reported the uncorrelated CPs detected in this analysis, together with the that measured, considering the same model for the CP, in \cite{Chen+2021}. Where the evidence in support of CP1 over TN is weak, the posterior distributions for $\log_{10}A_{CP}$ and $\gamma_{CP}$ are semi or unconstrained,
as can be noticed form the error bars representing the $68\%$
credible interval. As soon as the data contain from $5\%$ to $10\%$ of outliers, the presence of an uncorrelated CP becomes clear. The evolution of the
amplitude and the spectral index with the number of outliers resembles
those of the RNs of the pulsars or of the uncorrelated CP recovered in
Sec.~\ref{sec:Signals recovery with outliers} and Sec.~\ref{sec:Common process recovery}. Notably, the uncorrelated CP that
can be recovered when $1\%$ ($5\%$) of outliers is present in
the \textit{TwoF} (\textit{MultiF}) dataset, tends to overlaps with the CP recovered in the EPTA DR2 \citep{Chen+2021}.
\subsubsection{CP2 vs TN}
\begin{figure*}[ht] 
  \begin{minipage}  [b]{.33\textwidth}
    \centering
    \includegraphics[scale=0.43]{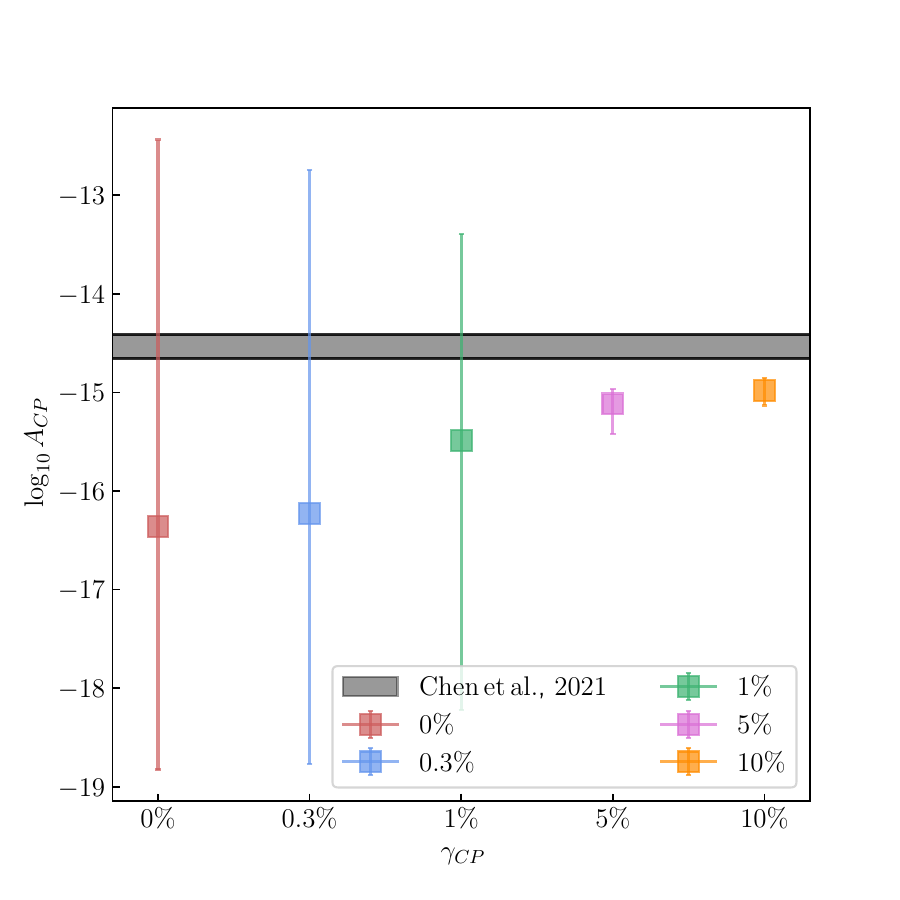}
  \end{minipage}%
  \begin{minipage}[b]{.33\textwidth}%
    \centering%
    \includegraphics[scale=0.43]{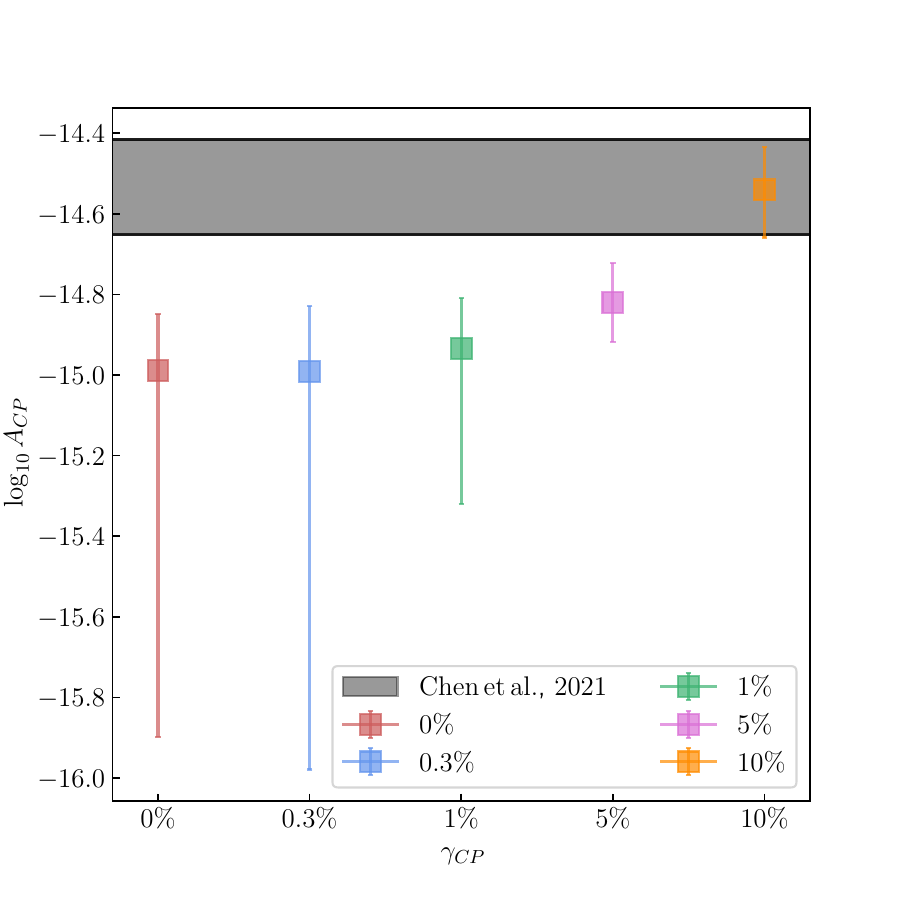}
  \end{minipage} 
  \begin{minipage}[b]{.33\textwidth}%
    \centering
    \includegraphics[scale=0.43]{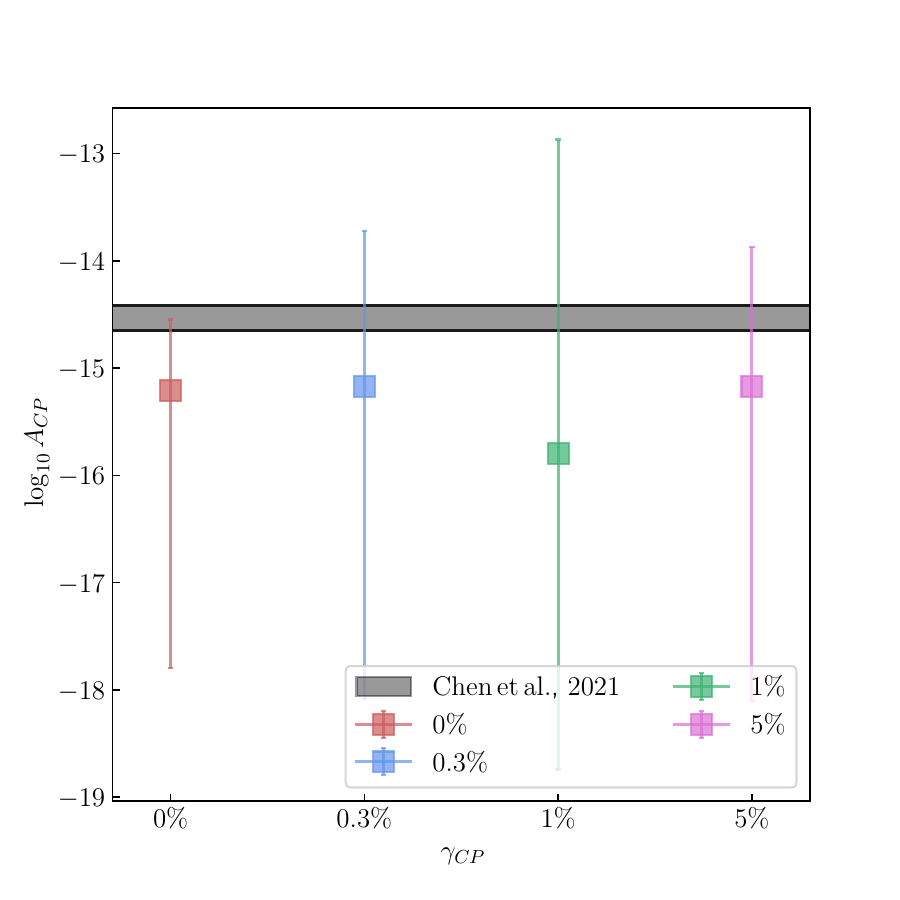}
  \end{minipage}%
  \caption{\small{The amplitudes of the CP found from the comparison between
    the models CP2 and TN for the datasets \textit{OneF} (left), 
    \textit{TwoF} (center) and \textit{MultiF} (right) dataset %
    corrupted with $0\%$ (red),  $0.3\%$ (blue), $1\%$ (green), $5\%$ (pink) and $10\%$ (orange) of outliers.  The error
    bars represent the 68\% credible interval. The black band represent the value of the amplitude recovered, modelling the CP as in CP2, in \cite{Chen+2021}. 
    }}
  \label{fig:CP2vsTN} 
\end{figure*}

We also observed increasing evidence in support of CP2 over TN, with the
growth being correlated with the percentage of outliers injected for
the \textit{OneF} and \textit{TwoF} %
datasets. According to Table \ref{tab:bf_CP2}, when in the data are present from $0$ to $1\%$ of outliers, there is weak and slowly growing support for CP2, and with $5\%$ and $10\%$ of outliers, this support becomes fairly substantial, in particular for the dataset \textit{TwoF}. The \textit{MultiF} dataset behaves slightly differently showing no evidence either in support of or against the CP2 model over TN. It can be noticed that the posterior odds ratios given in Table \ref{tab:bf_CP2} are orders of magnitude smaller than those reported in Table \ref{tab:bf_CP1}, especially those associated to the largest percentages of outliers, indicating that the CP2's support against TN is weaker than the CP1's. This agrees with the findings of Sec. \ref{sec:Common process recovery}. The uncorrelated CP found in there was characterized by  a relatively shallow slope, comparable to what found in
the comparison between the CP1 and TN models. Therefore, we expect that models that fix the spectral index at $13/3$ to  be less supported.
\begin{table}
  \centering
  \caption{\small{$\log _{10}$ posterior odds ratios obtained from the model comparison between CP2 and TN for the datasets studied for the percentages of outliers injected. For the
      \textit{MultiF} %
      dataset, this analysis in the scenario of $10\% $ of
      outliers injected in the data, has not been
      performed. Uncertainty over the final digit is indicated
      by the digit in parentheses. %
      }}
  	\resizebox{\columnwidth}{!}{
    \begin{tabular}{c*{6}{>{$}c<{$}}} 
            \hline
            Dataset& 0\%&0.3\%&1\%&5\%&10\%\\
            \hline
  \textit{OneF}&-0.316(5)&-0.231(5)&0.054(6)&0.87(1)&1.37(2)\\
  \textit{TwoF}&0.63(1)&0.608(9)&0.84(1)&2.2(1)&4.9\\
  \textit{MultiF}& 0.25(1)&0.135(9)&-0.171(6)&0.147(9)&-\\
  \hline
  \end{tabular}}
  \label{tab:bf_CP2}
\end{table}
The amplitudes of the CP recovered for these datasets are reported and compared with that found in \citet{Chen+2021}, while searching for the same kind of signal, in
Figure \ref{fig:CP2vsTN}. For the \textit{OneF} dataset
where the evidence in support of CP2 over TN is weak, $A_{CP}$ is
unconstrained. When
the percentage of outliers grow, so does the support for CP2 and
the posteriors become constrained and well defined. Conversely, even with no or a
few outliers, the recovered CP amplitude 
is always tightly constrained for the \textit{TwoF} dataset. The fact that a well-constrained
amplitude can be recovered even in the absence of outliers in the data
suggests, as already observed in Section \ref{sec:Common process recovery}, that some other property (other than outliers or GWB) of
the data may be culprit, making it difficult to
definitively pinpoint outliers as the primary cause of
CP in this dataset. However, it is clear that outliers have a significant impact on the CP when examining the strong change in amplitude as the number of outlier increases. Finally, despite the large number of outliers in the
\textit{MultiF} dataset, the recovered amplitude of the CP is
never constrained. We note that, in general, the amplitudes retrieved do not change
as dramatically as those of the CPs recovered from the comparison between CP1 and TN.

The majority of the values we recovered tend to overlap with that identified in \citet{Chen+2021}. Notably, our findings are especially relevant when considering the \textit{TwoF} dataset. When we introduce a $10\%$ outlier contamination into the dataset, the amplitude we recover closely matches that observed in real data. However, it is essential to emphasize that this type of analysis does not provide sufficient evidence to claim the detection of a GWB. Nevertheless, it is reasonable to conclude that outliers have clearly the potential to introduce a CP component comparable to what is observed in real data and could have contribute to it.
\section{DISCUSSION}\label{sec:DISCUSSION}
\subsection{The influence of outliers on signal
  recovery}\label{sec:Influence of outliers on signal recovery}

Based on the results presented in Section \ref{sec:RESULTS},
we found that for a GWB signal in the loud regime
($A_{GW} \gtrsim 2 \times 10^{-15}$) injected in an outliers-corrupted dataset, the recovered signal is always
well constrained and close to the injected value. %
However, even the smallest percentage of outliers caused a failure of RN and DM-induced noise parameters recovery. These three processes, which behave very similarly in the individual pulsar datasets since they all induce a time
(auto-)correlation between timing residuals, have one
significant difference: the GWB also induces a correlation between the timing residuals of different pulsars. Due to this propriety of the GWB, its recovery is largely unaffected by the presence of even a significant number of outliers (10\% of the data, in the worst case scenario we considered).

\subsubsection{The nature of the PTA covariance matrix and its
  implications on the signal recovery robustness.}\label{sec:covmat_disc}

The likelihood in \ref{eq:likelihood} depends on the product of
timing residuals and on the inverse of their covariance matrix
\citep{vanHaasteren+2009}. Specifically, the
products of timing residuals are divided by the corresponding elements
of the theoretically-calculated covariance matrix, and then summed together. This process is
iterated over different parameter values. Better timing parameter
estimates decrease this sum, maximizing the likelihood, while incorrect values decrease it.
Given the particular shape of the covariance matrix, we now show that the
most affected parameters are those lying on the diagonal part of the matrix.

Consider an $N \times N$ block matrix (see Figure \ref{fig:covmatrixdraw} for an example) where $N = \sum_{a=1}^{N_p} n_a$, and $a$ identifies a specific pulsar. Here, $N_p$ represents the number of pulsars in the array, and $n$ denotes the number of timing residuals per pulsar. Assume that there are $y\, n_a$ outliers for each pulsar, where $y$ is a percentage value ranging from 0 to 1. The number of permutations of $n $ distinct objects grouped $k$ at a time can be written as:
\begin{equation}\label{permutazioni}
 _nP^k=\frac{n!}{(n-k)!}.
\end{equation}
To evaluate the number of encounters of an outlier with the other timing
residuals, we set $k=2$, reducing Eq. \eqref{permutazioni} to $n(n-1)$.
Given a diagonal $n_a \times n_a$ block matrix with $y\,n_a$ outliers,
the number of intersections is then
$y\,n_a(y\,n_a-1)$. Comparing this number to the total number of possible
encounters ($n_a^2$) gives us the encounter density along the
diagonal of the covariance matrix of an array of $N_p$ pulsars:
\begin{equation}\label{eq:rhodiag}
 \rho_{\rm diag}=\sum_{a=1}^{N_p}\left(\frac{y\,n_a(y\,n_a-1)}{n_a^2}\right).
\end{equation}
The WN, RN, DM-induced noise and the auto-correlated part of the GWB all contribute to
this part of the matrix, as shown in Figure
\ref{fig:covmatrixdraw}. The density of encounters in the off-diagonal
parts of the covariance matrix, which corresponds to the cross-correlated
part of the GWB, is
\begin{equation}\label{eq:rhoff}
 \rho_{\rm off}=\frac{(\sum\limits_{a=1}^{N_p}
   y\,n_a-1)(\sum\limits_{a=1}^{N_p} y\,n_a)-\sum\limits_{a=1}^{N_p}
   (y\,n_a(y\,n_a-1))}{(\sum\limits_{a=1}^{N_p}
   n_a)^2-\sum\limits_{a=1}^{N_p} n_a^2}.
\end{equation}
If $y\,n_a \gg 1$, as we would expect for realistic PTA datasets like IPTA DR2, %
 the following approximations can be made: $\rho_{\rm diag}\sim
\sum_{a=1}^{N_p} y^2$ and $ \rho_{\rm off}\sim y^2$, which leads to:
\begin{equation}\label{finalresult}
 \frac{\rho_{\rm off}}{\rho_{\rm diag}}\sim \frac{1}{N_p}.
\end{equation}

Thus the most sensitive part of the covariance matrix is the diagonal,
where the encounter density is the highest.
The RN and DM-induced noise parameters, lying exclusively on that diagonal, are most strongly
affected compared to the off-diagonal dominated GWB parameters.
Notably, the density ratio of Eq.\eqref{finalresult} scales inversely with the number
of pulsars, showing that the GWB recovery is made more robust by adding more
pulsars.
\subsection{Outliers as sources of an uncorrelated CP}
Having determined the influence of outliers on the recovery of the
signals injected, we investigate outliers as a possible
source of common uncorrelated RN, as this can still contribute to the signal recently observed by the major PTA collaborations. 
We added to the model used in Section \ref{sec:Signals recovery with
  outliers} an uncorrelated CP modelled as a power-law characterized
by an amplitude $A_{CP}$ and a spectral index $\gamma_{CP}$ and then we
searched for all the other parameters (RN, DM, GWB) along with it. After testing our datasets to check if no uncorrelated CP could
be detected prior to the injection of the GWB or outliers, we
discovered that for the majority of the dataset, injecting the GWB
without outliers was sufficient to detect an uncorrelated CP. This
feature, as underlined in Section \ref{sec:Common process recovery},
could be related to some power coming from the auto-correlation part
of the GWB signal, which is detected as an uncorrelated CP. 
After adding outliers, the presence of a CP process became clear in each dataset studied.

In agreement with the findings in Section \ref{sec:Signals recovery with
  outliers} outliers do not influence the GWB recovery but clearly
affect that of the uncorrelated CP. %
In general, an evolution of
$A_{CP}$ toward larger values ($\sim 10^{-13}$) and of $\gamma_{CP}$ toward
the lower limit of the prior space ($\sim 0$) is readily seen, in each dataset, as the
number of outliers increases highlighting a correlation between such signal and outliers. However, it is
crucial to stress that, with the exception of the \textit{MultiF}
dataset, proper recovery of the GWB can always be achieved when the
latter is searched along with an uncorrelated CP. If a model without
an additional CP is utilized, the GWB can still be retrieved from the
\textit{MultiF} dataset despite outliers (see Section \ref{sec:Signals recovery with outliers}), and the
failure of the recovery during this analysis is only due to a "split"
of power between the uncorrelated CP and the GWB.%

To have a clear picture on the uncorrelated CP originating from outliers, %
we performed the models comparisons
presented in Section \ref{sec:Model selection}. From those, it is clear
that an uncorrelated CP can be measured if a high
enough percentage of outliers is present ($\geq 1\%$) in the data, and its proprieties are strictly related to their abundance.

 Figures \ref{fig:A_G_evoul_onef},\ref{fig:A_G_evoul_twof},\ref{fig:A_G_evoul_multif}
 and \ref{fig:A_G_DM_evoul_multi} show why we can detect a CP when
 outliers contaminate data. %
 \citet{Zic+2022} demonstrated that if the
 RNs of the pulsars share very similar amplitudes and spectral indices, it
 is possible to detect an uncorrelated CP from data that do not
 contain it. In particular this CP is characterized by
 an amplitude and spectral index that resemble those of the
 pulsars. As explained in Section \ref{sec:Influence of outliers on
   signal recovery}, the RNs and DMs are the processes that are most
 affected by outliers, which cause their amplitudes and spectral
 indices to tend to cluster, respectively, toward the upper and lower
 limits of their prior spaces. As a result, outliers are responsible for
 causing the pulsars to share very similar properties in terms of the
 RN and DM, thus leading to the recovery of a spurious CP. %
 Since the severity of the distortion in the RN/DM
 depends on the number of outliers, the characteristics of the CP
 recovered vary with it.

\section{CONCLUSIONS}\label{sec:CONCLUSION}

In light of the recent evidence for the GWB presented by PTAs \citep{eptadr2III,ng15yr,pptadr3,cptadr1}, we presented the first attempt at quantifying how much those results can be affected by the presence of bad data (i.e., outliers) in PTA data streams. To this end, we tried to answer the following three questions:
a) How can outliers influence the detection of the signals characterizing PTA data? b) Could outliers be the source of a CP? c) Could outlier-induced CP mimic the early appearance of a GWB?

To answer the first question, we considered a model that included WN
(kept fixed), RN, DM-induced noise, and the GWB, and we tried to recover the
GWB and all noise components injected in the data in the presence of outliers. The results of
this analysis, reported in Section \ref{sec:Signals recovery with
  outliers}, showed that the RN and DM-induced noise parameters were strongly affected by the smallest percentage of outliers, while the estimate of the GWB is robust
against any percentage injected, and this behavior can be deduced from the particular shape of the
likelihood used to perform parameter estimation.

To answer the second question, we added to the recovery model an uncorrelated CP modelled as
a power-law characterized by $A_{CP}$ and $\gamma_{CP}$. For all the datasets, an uncorrelated CP,
whose characteristics change with the number of outliers injected, is recovered.
This indicates that outliers can actually be a source of a spurious CP.
For some datasets, we also find that, as soon as the
GWB is injected into the data without outliers, an uncorrelated CP
was recovered alongside the GWB. In the most extreme case, which
has been observed when considering the \textit{MultiF} dataset, only an
uncorrelated CP could be recovered, burying the GWB (see Figure
\ref{fig:CP+GWB_multif}). The HD curve predicts that the
correlated component of the GWB is weaker than the uncorrelated
one; therefore, it is possible that some of the power of the GWB
has leaked into the uncorrelated CP, making it more challenging to
identify the GWB as a correlated process. This behavior was indeed
predicted by \citet{Romano+21,Pol+21}. They proposed that the GWB will
likely first appear as an uncorrelated CP before becoming a spatially
correlated signal as data gain enough sensitivity with time. We
increased the time period of the \textit{MultiF} dataset by $10$ years and found that this is indeed the case (see Figure \ref{fig:CP+GWB_multif+10y}). 

To answer the third question, we performed a model comparison
following \citet{Zic+2022}, considering data in which no GWB had
been injected but there were the contributions of WN, RN, DM-induced noise and
outliers. We examined models that included an
uncorrelated CP with a power-law shape and a variable or fixed
spectral index versus models that do not. We found strong support for
the models that include a CP, confirming once more that outliers can
be sources of such a signal and can potentially contribute to the uncorrelated component of the signal  recently observed.

These answers enabled us to draw the following important conclusions. When a GWB present in the data, no outliers can
successfully obscure or obliterate the signal (if the data are
sensible enough to recover it). Therefore, the pipelines currently used to analyze PTA data are robust against outliers when it comes to the characterization of a GWB. On the other hand, outliers can significantly
damage the RN's and DM-induced noise's detections to the point of producing an
uncorrelated CP. %
We found that such a signal, whose properties depend on the nature and quantity of outliers, can in some cases be compatible with, or at least contribute to, the uncorrelated RN component recently observed.

\section*{Acknowledgements}
We thank Aurélien Chalumeau and Joris Verbiest for discussions.
The authors acknowledge the support of colleagues in the EPTA. 
GF is supported by ERC Starting Grant No.~945155--GWmining, 
Cariplo Foundation Grant No.~2021-0555, MUR PRIN Grant No.~2022-Z9X4XS, 
and the ICSC National Research Centre funded by NextGenerationEU.  %
AS and GS acknowledge the financial support provided under the European Union’s H2020 ERC Consolidator Grant ``Binary Massive Black Hole Astrophysics" (B Massive, Grant Agreement: 818691). %

\bibliographystyle{aa} \bibliography{outliers}

\end{document}